\documentclass[%
 reprint,
superscriptaddress,
 amsmath,amssymb,
]{revtex4-2}

\usepackage{graphicx}
\usepackage{dcolumn}
\usepackage{bm}
\usepackage[hidelinks]{hyperref}
\usepackage{physics}
\usepackage{float}
\usepackage[dvipsnames]{xcolor}
\usepackage{tikz}
\usepackage{amsmath}
\usetikzlibrary{positioning, arrows.meta, backgrounds}

\usepackage{booktabs, colortbl, array, xcolor}
\renewcommand{\arraystretch}{2.2} 

\begin{document}


\title{Tensorial symmetries and optical lattice thermalization}

\author{Savvas S. Sardelis}
\affiliation{Mathematics Department, Florida State University, Tallahassee, Florida 32306, USA}
\author{Konstantinos G. Makris}
\affiliation{ITCP-Physics Department, University of Crete, Heraklion 71003, Greece}
\author{Ziad H. Musslimani}
\affiliation{Mathematics Department, Florida State University, Tallahassee, Florida 32306, USA}
\author{Demetrios N. Christodoulides}
\affiliation{Department of Electrical and Computer Engineering, University of Southern California, Los Angeles, CA
90089, USA}

\date{\today}
\begin{abstract}
This paper presents a comprehensive and systematic study of the possible connection between thermalization of cubic nonlinear lattices with nearest-neighbor coupling and the structure of the mixing tensor that arises due to the presence of nonlinearities. The approach is based on rewriting the underlying lattice system as a nonlinear evolution equation governing the dynamics of the modal amplitudes (or projection coefficients). In this formulation, the linear coupling become diagonalizable, whereas all cubic nonlinear terms transform into a combinatorial sum over a product of three modal amplitudes weighted by a fourth-order mixing tensor. The question that then arises is: can one extract any information regarding thermalization of the original lattice system solely from knowledge of the internal structure of the nonlinear wave mixing tensor? To this end, we have identified the exact structure of several mixing tensors (corresponding to different types of cubic nonlinearities) and tied their symmetry properties (quasi-Hermiticity and permutation symmetries associated with two lattice conservation laws) with thermalization or lack thereof. Furthermore, we have observed through direct numerical simulations that the modal occupancies of lattices preserving these tensorial symmetries approach a Rayleigh-Jeans distribution at thermal equilibrium. In addition, we provided few examples that indicate that cubic lattices with broken tensorial symmetries end up not to equilibrate to a Rayleigh-Jeans distribution. Finally, an inverse approach to the study of thermalization of cubic nonlinear lattices is developed. It establishes a duality property between lattices in local and modal bases. The idea is to establish a trade-off between the type of nonlinearities in local base and their respective interactions in supermode base. With this at hand, we were able to identify a large class of nonlinear lattices that are embedded in the modal space and admit a simple form that can be used to shed more light on the role that localization (or delocalization) of the supermodes play in thermalization processes. 
\end{abstract}
\maketitle
\section{\label{sec1}Introduction}
In recent years, considerable interest has emerged in the study of complex physical systems characterized by many interacting degrees of freedom. A paradigm example is nonlinear wave propagation in multimode photonic waveguides \cite{poletti2008description,mafi2012pulse,renningerWise2013,richardson2013spdivmultiplex,wright2015controllable,liu2016kerr,lopez2016visible,krupa2016spatiotemporal,krupa2016observation,krupa2017spatial,fusaro2019dramatic}. In such settings, the governing coupled equations of the corresponding optical amplitudes are of Schr\"odinger type in the presence of nonlinearities that account for multiwave mixing. In this regard, the main underlying challenge is rooted in the computational complexity of such highly nonlinear multimode processes. Recently, a novel approach based on statistical mechanics has been developed to understand the behavior of complex photonic systems \cite{FanHassan,MidyaFanPawel}. Rather than monitoring the dynamics of each individual mode-a computationally expensive task-the theory of optical thermodynamics instead provides statistical information about the collective behavior of all interacting modes at thermal equilibrium \cite{Makris,FanPawel2020,Efremidis,Kottos2021,ramos2023Kottos,Ren_Pressure,Efremidis_pressure}. Its success in accurately predicting the relaxation states of complex nonlinear multimode photonic structures has also been confirmed in recent experiments \cite{Pourbeyram2021, pourbeyram2022, wu2022mode-locked,wright2022physics,ferraro2023review, ManginiFerraro,ferraro2024calorimetry}. A simple and illustrative example can be provided by the discrete nonlinear Schr\"odinger (DNLS) equation  with nearest neighbor coupling and Kerr nonlinearity \cite{ChristodoulidesJoseph,ChristodoulidesSilberberg}. It has been shown that when starting with a random sample of input optical fields, their modal occupancies (i.e., optical power per mode) relax to a Rayleigh-Jeans (RJ) distribution after extended propagation \cite{selim2022Near_Far}. Beyond the DNLS regime, the Rayleigh-Jeans law has been linked to irregular power distributions of non-Hermitian lattices \cite{pyrialakos2022PT}, and a generalized RJ distribution has also been derived to incorporate the conservation of orbital angular momentum in cylindrical multimode nonlinear optical waveguides \cite{wu2022angular}. Recently, the theory of optical thermodynamics has also been applied to investigate the equilibrium behavior of nonlinear topological optical systems \cite{jung2022topological}. The presence of Kerr nonlinearity in all these examples facilitates mode interactions and through ergodic processes the system steers towards its equilibrium state \cite{Kottos2011, Kottos2020}. Notably, it has been argued that the process of thermalization is independent of the specific form of the nonlinearity \cite{Zhong2023}. Nonetheless, while it constitutes a necessary ingredient for the system to attain an equilibrium state, its role is not fully understood. Recent work highlights the intricate role that nonlinearity plays within the context of integrable models \cite{baldovin2021integrable}. Specifically, it was revealed through numerical simulations that the integrable Ablowitz-Ladik (AL) model fails to thermalize \cite{SelimPyrialakos_Ablowitz}.
In this paper, we study the thermalization properties of several nonlocal optical lattice models within the framework of the DNLS equation \cite{rasmussenKevrekidisStatistical,kevrekidisDiscreteBook, ablowitz2002discrete, ablowitz2003discrete}. In this context, thermalization denotes the statistical equilibrium state (sort of a nonlinear attractor) of the governing equation, to which a wide range of random initial conditions converges upon averaging. In particular, we investigate the possibility that the symmetries of the nonlinear mixing tensors, associated with power and energy conservation, are intimately linked to the thermal equilibrium of cubic optical lattices.
We provide an in-depth exploration of the symmetry structure of these tensors in both integrable and non-integrable cubic lattices, aiming to shed further light on this relatively unexplored area.
Importantly, we develop an inverse approach in which optical lattice models are constructed in modal space and their thermalization properties are subsequently investigated.
The advantage of this methodology is that it allows one to derive simple models in the modal base that conserve power and energy, thereby facilitating an exploration of the role that nonlinear nonlocality (from short to long range) may play in the formation of an equilibrium state. 

This paper is organized as follows. In Section \ref{sec2}, we introduce the general family of nonlinear cubic equations that form the core of our study on both the local and supermode bases, along with the associated linear eigenvalue problem and their Hamiltonian structure. We also introduce two important tensorial symmetries that underpin our thermalization conjecture. In Section \ref{sec3}, we present a brief overview of the theory of optical thermodynamics. Section \ref{sec4}, introduces a specific family of cubic nonlinear lattices in the local basis. Section \ref{sec5} is devoted to the tensorial structure of each individual nonlinear component. In Section \ref{sec6}, we present numerical simulations designed to test the possible connection between nonlinear tensorial symmetries and optical thermalization. Additionally, we explore the dependence of the tensor on the supermodes and the flexibility of the evolution equation in the supermode base. Lastly, in Section \ref{sec7}, we present examples of cubic nonlinear lattices that do not conform to the tensorial symmetries and demonstrate how they fail to attain a Rayleigh-Jeans equilibrium distribution. 
\section{\label{sec2} Governing Equations}
\subsection{\label{sec2_1}Dynamics in local base}
We begin the discussion by considering a general framework upon which the current work is based. Specifically, the physical model under study is a finite size one-dimensional generalized 
DNLS type system that includes a nearest-neighbor coupling, on-site potential, and a collection of cubic nonlinearities given by
\begin{equation}
\label{generic-DNLS}
i\frac{dA_n}{dz} + A_{n-1}+A_{n+1} + V_nA_n + f(A_n,A_{n\pm m}) = 0\,,
\end{equation}
where $f(A_n,A_{n\pm \tilde n})$ is a complex cubic polynomial of the optical field amplitude $A_n$. Here, $n = 1,2,\dots, M$ indicates the lattice site, $z$ is the propagation distance, $V_n$ is the on-site potential and $\tilde n$ is an arbitrary integer. It is assumed that Eq.~(\ref{generic-DNLS}) 
remains invariant under the gauge transformation:
\begin{equation}
\label{gauge}
A_n \rightarrow A_ne^{i\theta} \;,
\end{equation}
with real constant $\theta$. While at this point the specific structure of the cubic nonlinearities appearing in Eq.~(\ref{generic-DNLS}) is left arbitrary, the gauge symmetry (\ref{gauge}) would eventually limit their form. In this regard, we have  
\begin{equation}
   f(A_n,A_{n\pm \tilde n}) \sim  A_{n+n_1}A_{n+n_2}A^*_{n+n_3}\, ,
    \label{cubicTerms}
\end{equation} 
where star indicates complex conjugation and $n_1$, $n_2$, $n_3$ are arbitrary integers. 
This means that lattices with nonlinearities of the form $|A_{n\pm n_1}||A_{n\pm n_2}|A_{n\pm n_3}$ (or their alike) are not included in our study.
\subsection{Eigenvalue problem}
\label{sec2_2}
In the linear regime (where all nonlinear cubic terms are absent) the ansatz 
\begin{equation}
\label{An}
A_n(z) = \psi_n e^{i\epsilon z} \;,
\end{equation}
leads to the following linear matrix eigenvalue problem:
\begin{equation}
 \mathbb{M} | \psi^{(j)} \rangle = \epsilon_j  | \psi^{(j)} \rangle \;,
    \label{eqEigen}
\end{equation}
where $\mathbb{M}$ is an $M\times M$ tridiagonal matrix whose main diagonal is the on-site potential $V_n$ and its two off diagonals are ones. The notation $| \psi^{(j)} \rangle,\,\,  j = 1,2, \dots, M,$ is used to denote the $j^{\text{th}}$ supermode, defined by
\renewcommand{\arraystretch}{0.8}
\begin{equation}
    | \psi^{(j)}\rangle 
     \equiv \{\psi_n^{(j)}\}_{n=1}^M
     =
    \begin{pmatrix}
        \psi^{(j)}_{1}\\
        \vspace{.01in} \\
        \psi^{(j)}_{2}\\
        \vdots\\
        \psi^{(j)}_{M}
    \end{pmatrix},
    \label{Supermodes}
\end{equation}
and $\epsilon_j$ being the matrix eigenvalues. The supermodes constitute an orthonormal base satisfying the orthogonality condition:
\begin{equation}
   \langle  \psi^{(j)} |  \psi^{(j')} \rangle 
 =\delta_{j,j'} \;.
\end{equation}
Without loss of generality, it is assumed that the eigenvalues are arranged in ascending order $\epsilon_1\leq\epsilon_2\leq \cdots \leq\epsilon_M$ where $\epsilon_1$ ($\epsilon_M$) corresponds to the highest (lowest) order supermode. For the special case where $V_n=0$ and in the presence of zero boundary conditions, i.e. $\psi^{(j)}_0=\psi^{(j)}_{M+1}=0$, an analytical form for the supermodes $\psi_{n}^{(j)}$ and the eigenvalues $\epsilon_j$ of Eq.\eqref{eqEigen} can be obtained. They are given by
\begin{equation}
    \epsilon_j=2 \cos(\frac{j\pi}{M+1})\, ,
    \label{RegLatEigval}
\end{equation}
\begin{equation}
    \psi_n^{(j)}=\sqrt{\frac{2}{M+1}}\sin(\frac{jn\pi}{M+1}) \, .
    \label{RegLatEigvec}
\end{equation}
\subsection{\label{sec2_3}General formulation in modal space}
We seek solutions to Eq.~(\ref{generic-DNLS}) in the form of
\begin{equation}
    A_n(z)=\sum\limits_{j=1}^M c_j(z)\psi_n^{(j)}\, ,
    \label{project1}
\end{equation}
where $c_j(z)$ are the so-called complex projection coefficients. Substituting the expansion (\ref{project1}) into (\ref{generic-DNLS}) we obtain the following generic nonlinear evolution equation governing the dynamics of the projection coefficients (or modal amplitudes): 
\begin{equation}
    i\frac{dc_j}{dz}+\epsilon_j c_j+\sum_{k,l,m=1}^{M} T_{j,k,l,m}\,c_k c_l c^*_m=0\, ,
    \label{SuperModeDNLSE}
\end{equation}
for all $j=1,2,\dots, M$. In the above, $T_{j,k,l,m}$ is a rank 4 (in general) complex mixing tensor that arises due to the presence of cubic nonlinear terms. For the type of nonlinearity shown in Eq.~\eqref{cubicTerms} this tensor takes the form
\begin{equation}
    T^{\text{typical}}_{j,k,l,m} =\sum_{n=1}^M\psi_n^{{(j)}^*}\psi^{(k)}_{n+n_1} \psi^{(l)}_{n+n_2} \psi^{{(m)}^*}_{n+n_3}\,.
    \label{tensorExample}
\end{equation}
Note that Eq.~(\ref{SuperModeDNLSE}) is invariant under the phase transformation
$c_j \rightarrow c_je^{i\vartheta}$ with real constant $\vartheta$. If one relaxes condition (\ref{gauge}) then three other groups of nonlinear cubic terms would appear. They assume the form:
$T_{j,k,l,m}\,c_k c^*_l c^*_m,\; 
T_{j,k,l,m}\,c^*_k c^*_l c^*_m$ and $T_{j,k,l,m}\,c_k c_l c_m.$ 
Such scenario is not considered in this paper and is left for a future study.
System (\ref{SuperModeDNLSE}) constitute the basis of our work as it provides a unified framework for the study of the statistical properties of DNLS systems with the aforementioned type of cubic nonlinearities. In this regard, the quantity of interest is defined by
\begin{equation}
   \langle |c_j|^2 \rangle\equiv \langle \lim_{z\to \infty}|c_j(z)|^2 \rangle\, ,
   \label{QuantityOfInterest}
\end{equation}
where $\langle \cdots \rangle$ denotes the ensemble average over many realizations of the random initial conditions $c_j(0)$. Of particular importance is the dependence of $\langle |c_j|^2 \rangle$ (at thermal equilibrium) on the eigenvalues $\epsilon_j$. We note that if the dynamical system admits a thermal equilibrium, then the associated quantity $\langle |c_j|^2 \rangle$ should be independent of the initial conditions $c_j(0)$. When this is the case, we have
\begin{equation}
    \langle |c_j|^2 \rangle=h(\epsilon_j)\, ,
\end{equation}
where $h$ denotes the energy distribution function. According to the theory of optical thermodynamics (see Sec.~\ref{sec3}) this functional dependence (for lattices with two conservation laws, i.e. power and energy) take the Rayleigh-Jeans form: 
\begin{equation}
    h=\frac{a}{\epsilon_j+b} \hspace{0.1cm}\, ,
    \label{RJD_1}
\end{equation}
where $a$ and $b$ are related to the conservation laws.

Throughout the rest of the paper, equation \eqref{SuperModeDNLSE} will be augmented with initial condition $c_j(0)$ and zero boundary conditions: $c_0(z)=c_{M+1}(z)=0$. The initial condition is prepared as follows: 
\begin{equation}
    c_j(0)\equiv \tilde c_j e^{2\pi i r_j}\, ,
\end{equation}
where $\tilde c_j$ is a deterministic positive real amplitude while the phase $r_j$ is a random real number sampled from a uniform distribution on the interval $(0,1)$. Furthermore, the amplitude $\tilde c_j$ is chosen to be one of the following (depending on the case at hand):
\begin{itemize}
    \item linear power distribution across all supermodes:
    \begin{equation}
        \tilde c_j=\sqrt{\alpha(\epsilon_j-\epsilon_1)}, \quad \alpha=\frac{P}{\sum_i  (\epsilon_i-\epsilon_1)}\, ,
        \label{LinearDistribution}
    \end{equation}
    \item piecewise uniform distribution across the supermodes:
    \begin{equation}
        \tilde c_j=
        \begin{cases}
            \displaystyle \sqrt{\frac{P}{j_2-j_1}} , \quad j_1\leq j\leq j_2\, ,\\
            \\
            0 , \quad \quad \quad \,\,\,\, \text{otherwise}\, .
        \end{cases}
        \label{EquipartitionDistribution}
    \end{equation}
\end{itemize}
Both cases ensure that the total initial power (at $z=0$), and consequently at every $z$, attains the predetermined value $P$ given by 
\begin{equation}
    P=\sum\limits_{j=1}^M |c_j|^2\, .
    \label{PowerSupermode}
\end{equation}
In what follows, we list some important assumptions about the symmetries of the tensor $T_{j,k,l,m}$.
\begin{align}
    &\text{Quasi-Hermiticity:}\,\, T_{j,k,l,m} = T^*_{l,m,j,k}\, , \label{QuasiHermiticity} \\
    \nonumber\\
    &\text{Permutation symmetry:}\,\, T_{j,k,l,m} = T_{m,l,k,j}\, .
    \label{Permutation}
\end{align}
Symmetry condition \eqref{QuasiHermiticity} guarantees the conservation of power. The second property \eqref{Permutation} allows one to derive Eq.~(\ref{SuperModeDNLSE}) from Hamilton's equations of motion for the conjugate pair of canonical variables $c_j,\,c_j^*$:
\begin{equation}
    \frac{dc_j}{dz}=i\{c_j,\mathcal{H}\},\;\;  \mathcal{H}=\mathcal{H}_0+\mathcal{H}_{\text{NL}}\, ,
    \label{HamiltonEq}
\end{equation}
where
\begin{equation}
    \mathcal{H}_0 =\sum\limits_{j=1}^{M}\epsilon_j|c_j|^2\, ,
    \label{LinearHamSupermode}
\end{equation}
and
\begin{align}
    \mathcal{H}_{\text{NL}}=
    \frac{1}{4}\sum\limits_{jklm}\left(T_{j,k,l,m} c^*_k c^*_l c_m c_j + T^*_{j,k,l,m}c_k c_l c^*_m c^*_j\right)\, .
    \label{NonLinearHamSupermode}
\end{align}
In Eq.~(\ref{HamiltonEq}), $\{\}$ denotes the standard Poisson bracket defined by
\begin{equation}
    \{D,\tilde D\}=\sum_{n=1}^M\left(\frac{\partial D}{\partial c_n}\frac{\partial \tilde D}{\partial c^*_n}
    - \frac{\partial D}{\partial c^*_n}\frac{\partial \tilde D}{\partial c_n}\right) \;,
    \label{PoissonBrac}
\end{equation}
where $D$ and $\tilde D$ are two arbitrary functionals of the canonical variables $c_n$ and $c^*_n$. Following this definition, one can derive the Poisson brackets for the respective canonical coordinates of interest, that is, $\{c_j^*,c_n\}=\delta_{j,n}$ and $\{c_j,c_n\}=\{c_j^*,c_n^*\}=0$. We reiterate that, in order to derive Eq.\eqref{SuperModeDNLSE} using the above Hamiltonian formulation, it is necessary to ensure that the tensor $T_{j,k,l,m}$ remains invariant under the permutation symmetries
$k\leftrightarrow l$ and $j\leftrightarrow m$. A detailed derivation of these two symmetries is presented in Appendix A.
\section{\label{sec3} Optical thermodynamics: An Overview}

As mentioned in the introduction, one of the main goals of this paper is to examine the concept of thermalization from the viewpoint of the symmetry properties of the underlying nonlinear mixing tensors. This will be done in conjunction with the newly developed theory of optical thermodynamics \cite{FanHassan,MidyaFanPawel,Makris}. In this section, we review the basic concepts and ideas of optical thermodynamics as they fit into our theoretical framework. The main result of the optical thermodynamics theory is that a weakly nonlinearly coupled optical system which conserves power and the full Hamiltonian will, upon reaching thermal equilibrium, exhibit a power partition across its modes that conforms to a Rayleigh-Jeans (RJ) distribution. This equilibrium state can be determined exclusively by the problem's linear spectrum and total power. More specifically, from the modal space point of view, as previously discussed, the power is given by Eq.~\eqref{PowerSupermode} whereas the internal energy of the system is defined by
\begin{equation}
    U\equiv-\mathcal{H}_0=-\sum\limits_{j=1}^{M}\epsilon_j|c_j|^2\, .
    \label{internal_Energy}
\end{equation}
Since the optical system is weakly nonlinear, this implies that most of its energy is stored into $U$ while a small portion is transferred to the nonlinear coupling. When this is the case, we can think of the internal energy given by Eq.~\eqref{internal_Energy} as being a constant of motion. Additionally, thermal equilibrium is attained by the system through the maximization of its entropy \cite{ManginiFerraro}, which, for the discussed nonlinear waveguide lattices, leads to a RJ distribution for the expected values of the modal occupancies
\begin{equation}
    \langle |c_j|^2 \rangle=-\frac{T}{\epsilon_j+\mu} \, ,
    \label{RJD}
\end{equation}
where $T$ is the optical temperature and $\mu$ the chemical potential \cite{Picozzi1,Picozzi2,DYACHENKO,FanHassan,Makris,MidyaFanPawel}.
\section{\label{sec4}Family of cubic optical lattices: local base formulation}
So far, we have discussed generic cubic lattices alongside their properties in modal space. In this section, we focus the attention on specific cubic lattices within the context of DNLS equation in the presence of various local/nonlocal cubic nonlinearities. We start with the Hamiltonian $H=H_0+H_{\text{NL}}$, where $H_0$ corresponds to nearest-neighbor coupling and on-site energy potential defined by
\begin{equation}
    H_0 =\sum\limits_{n=1}^M\Big(A_{n}^*A_{n+1}+A_{n+1}^* A_{n} + V_n|A_n|^2\Big)\, .
    \label{Hamilt2}
\end{equation}
The second term in the Hamiltonian, assumes the form
\begin{equation}
    H_{\text{NL}}= H_1+H_2+H_3\, ,
    \label{NonLinearHam}
\end{equation}
where 
\begin{equation}
    H_1 = \frac{g}{2}\sum\limits_{n=1}^M|A_n|^4\, ,
    \label{KerrNonLin}
\end{equation}
corresponds to the Kerr nonlinearity with $g$ being a real constant. Furthermore,
\begin{equation}
    H_2 =\frac{1}{2}\sum\limits_{n=1}^M\left(a_1|A_n|^2|A_{n+1}|^2+a_2 A_n^2(A_{n+1}^*)^2\right)+\text{c.c.}\, ,
    \label{NonLocalNonLin}
\end{equation}
with real coupling constants $a_1$ and $a_2$ and $\text{c.c.}$ stands for complex conjugation. Each individual term in the Hamiltonian induces a lattice with cubic nonlocality. Lastly,
\begin{equation}
    H_3 = \sum\limits_{n=1}^M   A_n^*A^*_{n+1} \left(b_1  A_n^2  
    + b_2  A^2_{n+1}  \right) + \text{c.c.}\, ,
    \label{ExtALFullLatNonLin}
\end{equation}
where $b_1$ and $b_2$ are real parameters. In the case where $b_1=b_2$ the resulting cubic nonlinear lattice embodies the Ablowitz-Ladik \cite{ablowitzLadik1976,ablowitz1991solitons} as well as a variety of other cubic nonlocal nonlinear terms \cite{oster2003FullLattice,KevrekidisKonotop_FullLattice}. 
Using Hamilton's equation of motion for the canonical variables $A_n$ and $A_n^*$
\begin{equation}
    \frac{dA_n}{dz} = i\{A_n,H\}\, ,
    \label{HamiltonEq2}
\end{equation}
where $\{\,\}$ represents the standard Poisson bracket, we arrive at the following dynamical system:
\begin{eqnarray}
    -i\frac{dA_n}{dz}&=& A_{n-1}+A_{n+1} +V_nA_n+g|A_n|^2A_n \nonumber\\
    \nonumber\\
    &+&a_1A_n\left(|A_{n-1}|^2+|A_{n+1}|^2\right)\nonumber\\
    \nonumber\\
    &+&a_2A_n^*\left(A_{n-1}^2+A_{n+1}^2\right)\nonumber\\
    \nonumber\\
    &+& b_1\left(2|A_n|^2A_{n+1}+A_n^2A_{n+1}^* + |A_{n-1}|^2A_{n-1}\right)\nonumber\\
    \nonumber\\
    &+&b_2\left(2|A_n|^2A_{n-1}+A_n^2A_{n-1}^* + |A_{n+1}|^2A_{n+1}\right)
    \, ,
    \nonumber\\
    \label{FullLattice}
\end{eqnarray}
for $n=1,2,\dots,M$. Here, $g$, $a_j$, $b_j$, $j=1,2$ are real parameters that assume the values of 0 or 1 depending on the lattice under consideration. Equation \eqref{FullLattice} exhibits two conserved quantities: the total energy, encapsulated by the Hamiltonian $H$ (Eq.~\eqref{NonLinearHam}) and the total power
\begin{equation}
    P=\sum\limits_{n=1}^M|A_n|^2.
    \label{Power}
\end{equation}
The connection between the system in Eq.~\eqref{FullLattice} and its modal counterpart (Eq.~\eqref{SuperModeDNLSE}), can be established by utilizing the eigenfunction expansion given in Eq.~\eqref{project1}.
\section{\label{sec5} Lattice Tensors: Tree structures and exact representations}
As mentioned earlier, one of our main goals in this paper is to establish a possible connection between thermalization of nonlinear cubic lattices and their associated mixing tensor. The mixing tensor arises solely due to the presence of nonlinear interactions which helps the system reach equilibrium (when it exists). Furthermore, the tensorial structure is fully determined by the properties of the non-interacting system, i.e., the eigenenergies and their respective supermodes, as well as the nature of the nonlinearity. As such, understanding the properties of tensors for lattices that reach thermal equilibrium (as well as those that do not) would shed light on the role that nonlinear interactions play in reaching thermalization (or its lack thereof). With this in mind, in this section, we provide a closed form expression for the tensor $T_{j,k,l,m}$ that appears in Eq.~\eqref{SuperModeDNLSE} corresponding to the interaction terms mentioned in Sec.~\ref{sec4}. These findings will later be tied to thermalization. In what follows, we list three major advantages of the tensor analysis approach. These are
\begin{itemize}
    \item The tensor for any fixed cubic lattice (subject to the symmetry constraints that we imposed earlier) will always be expressed as a combinatorial sum over a product of four supermodes (see Eq.~\eqref{tensorExample}). In other words, the tensor architecture is directly associated with the eigenvectors of the linear problem.
    \item The presence of multiple cubic terms (in local base) would alter the functional form of the discrete four-dimensional mixing tensor while leaving the structure of the dynamical system Eq.~\eqref{SuperModeDNLSE} unchanged. 
    \item The values of the mixing tensor uniquely determine the nonlinear coupling weights and the various combinatorial terms $c_k c_l c^*_m$.
\end{itemize}
Thus, the mixing tensor encodes the information properties inherited from both the linear and nonlinear problems. This highlights the importance of its study in conjunction with thermalization. In this paper and for the sake of simplicity, we restrict our analytical study to the special case of free lattices ($V_n=0$) for which a closed form of the supermodes is given by Eq.~\eqref{RegLatEigvec}. Note that for periodic potentials satisfying $V_{n+M}=V_n$ one can use Floquet theory to represent the supermodes as a Fourier series. This approach brings resemblance to Eq.~\eqref{RegLatEigvec}. In Section \ref{sec6}, we will comment on how tensorial symmetries and thermalization properties get affected in the presence of disordered potentials.
\subsection{\label{sec5_1}Kerr lattice}
\begin{figure*}[!hbt]
    \centering
    \includegraphics[width=\linewidth]{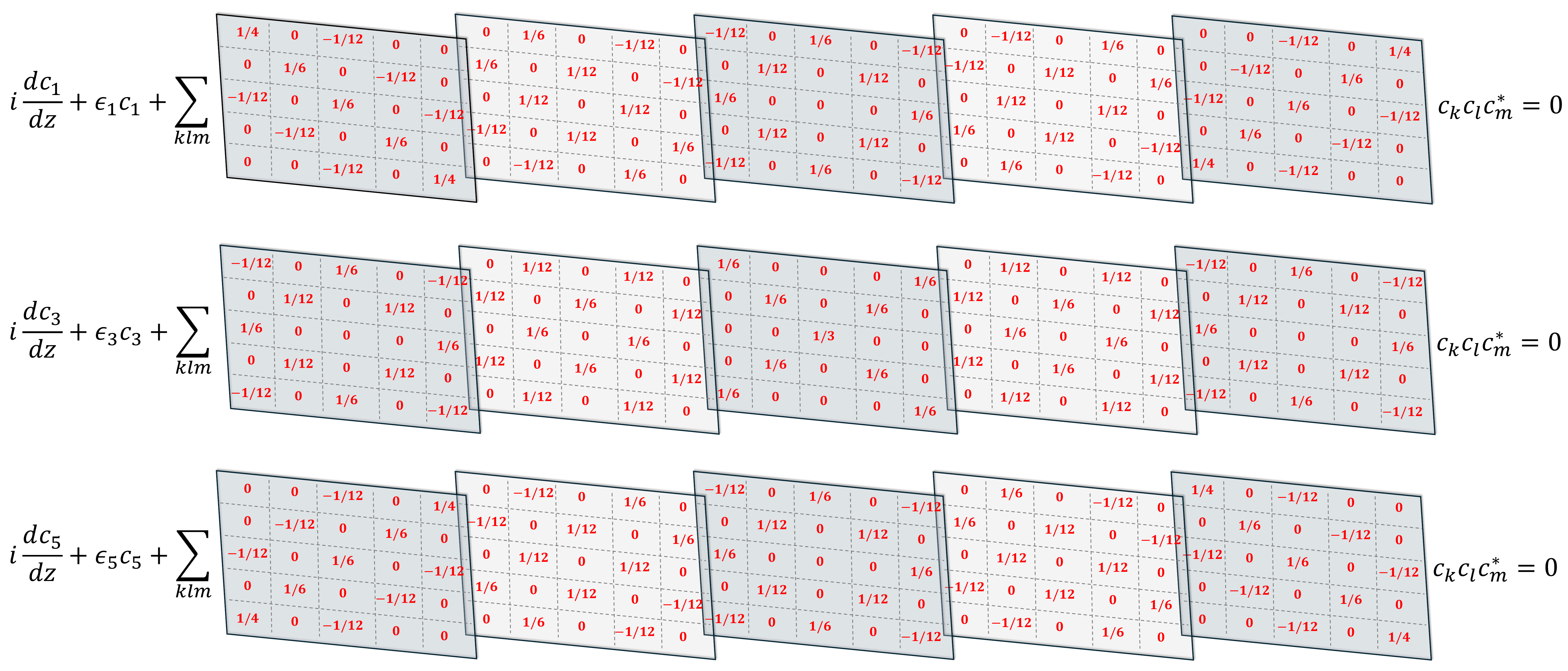}
    \caption{Evolution equations of the projection coefficients $c_j$ (for simplicity we only shown the $j=1,3,5$ elements) as given by Eq.~\eqref{SuperModeDNLSE} when the number of supermodes is $M=5$. Here, the tensor $T_{jklm}$ appearing in Eq.\eqref{SuperModeDNLSE} corresponds to the Kerr lattice as defined by Eq.~\eqref{KerrExact2}. The index $k$ counts the number of tensor slices oriented from left to right. For a fixed tensor slice the indices $l$ and $m$ denote the number of its rows and columns respectively.}
    \label{TensorKerr_M5}
\end{figure*}

Here, we will provide a closed form expression for the tensor $T_{j,k,l,m}\equiv T^{\text{Kerr}}_{j,k,l,m}$ in the presence of Kerr nonlinearity. To do so, we start from Eq.~\eqref{FullLattice} with $g=1$ while setting the rest of the parameters to zero. This leads to
\begin{equation}
    i\frac{dA_n}{dz}+ A_{n-1}+A_{n+1} +|A_n|^2A_n=0\, .
\end{equation}
Following similar ideas outlined in Sec.~\ref{sec2_3} (see also Appendix B for further details) we arrive at
\begin{equation}
    T^{\text{Kerr}}_{j,k,l,m} =  B\sum\limits_{n=1}^{M} \prod_{i=1}^4 \sin(q_i x_n)\equiv \frac{B}{16}\sum_{\iota=1}^8 (-1)^{\iota+1}\gamma_{j,k,l,m}^{(\iota)}\, ,
    \label{TensorKerrSine}
\end{equation}
where $B=4/(M+1)^2$ and $x_n=n\pi/(M+1)$. For the rest of the paper we will make a frequent use of the short set-type notation
\begin{equation}
    \{q_i\}_{i=1}^4 \equiv \{ j,k,l,m \},\, \, \,\,  \{p_i\}_{i=1}^4 \equiv \{ j,m,k,l \}.
    \label{IndicesPiQi}
\end{equation}
\begin{figure*}[!hbt]
    \centering
    \includegraphics[width=\linewidth]{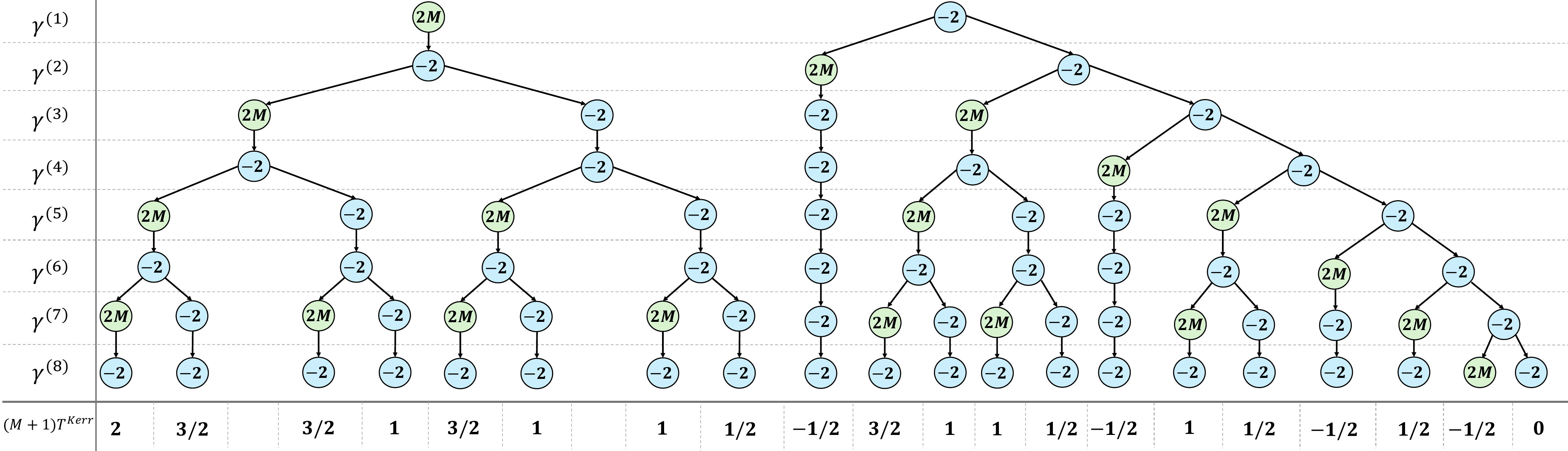}
    \caption{The tree diagram associated with the Kerr tensor depicting all possible branches that lead to the set $S$. The labels appearing on the far left vertical line represent the values of the auxiliary tensor $\gamma^{(\iota)}$ defined in Eq.~\eqref{KerrExactComponents}. For the ease of representation, the tensor indices have been suppressed. The horizontal axis corresponds to the values of the tensor for each respective tree branch as defined by Eq.~\eqref{TensorKerrSine}. The left tree starts form the ``root" $\gamma^{(1)}=2M$ while the right one from $\gamma^{(1)}=-2$. These ``roots" originate from the possibilities of $w_1$ precisely equals to $2M+2$ or an arbitrary even integer not equal to $2M+2$ as dictated by its bounds (see Table~\ref{Table_w}). An example of a tree branch is given by : $2M\to-2\to2M\to-2\to2M\to-2\to-2\to-2$, which leads to tensor value $T^{\text{Kerr}}=3/(2M+2)$. For a fixed set of indices $j,k,l,m$ one can uniquely identify any specific branch on the tree. The branches are labeled from left to right with the first branch appearing on the far left and the 20th branch on the far right.}
    \label{Tree}
\end{figure*}
The auxilary tensor $\gamma_{j,k,l,m}^{(\iota)}$ is given by
\renewcommand{\arraystretch}{1.5}
\begin{equation}
\gamma_{j,k,l,m}^{(\iota)} = 
\Bigg\{
    \begin{array}{lr}
        (-1)^{w_\iota+1}-1, & \text{if } w_\iota \neq 2(M+1)\kappa_\iota\, ,\\
        2M, & \text{if } w_\iota=2(M+1)\kappa_\iota\, ,
    \end{array}
    \label{KerrExactComponents}
\end{equation}
where $\kappa_{\iota}$ is an arbitrary integer and $w_\iota$ assumes one of the eight distinct integer combinations of the indices $j, k, l, m \in \{1,2,..,M\}$ presented in Table~\ref{Table_w}. 
\renewcommand{\arraystretch}{2.2}
\rowcolors{1}{blue!5}{gray!5}
\begin{table}[ht]
\centering
\begin{tabular}{>{\columncolor[gray]{0.9}}c|c}
\toprule
\rowcolor[gray]{0.92} \textbf{$\boldsymbol{w_{\iota}}$} & \textbf{Bounds} \\ \midrule
$w_1 = j + k + l + m$ & $4 \leq w_1 \leq 4M$ \\ 
$w_2 = j - k + l + m$ & $3 - M \leq w_2 \leq 3M - 1$ \\ 
$w_3 = -j - k + l + m$ & $2 - 2M \leq w_3 \leq 2M - 2$ \\ 
$w_4 = -j + k + l + m$ & $3 - M \leq w_4 \leq 3M - 1$ \\ 
$w_5 = j - k - l + m$ & $2 - 2M \leq w_5 \leq 2M - 2$ \\ 
$w_6 = j + k - l + m$ & $3 - M \leq w_6 \leq 3M - 1$ \\ 
$w_7 = -j + k - l + m$ & $2 - 2M \leq w_7 \leq 2M - 2$ \\ 
$w_8 = -j - k - l + m$ & $3 - M \leq w_8 \leq 3M - 1$ \\ 
\bottomrule
\end{tabular}
\caption{A list of all possible combinations of $w_{\iota}$, $\iota=1,2,\dots,8$ that arise from the derivation of Eq.~\eqref{TensorKerrSine} along with their upper and lower bounds (which helps determine if $w_{\iota}$ is equal to an integer multiple of $2M+2$ or not). These possibilities determine the ultimate value of the auxiliary tensor $\gamma_{j,k,l,m}^{(\iota)}$ and, in turn, the tensor $T_{j,k,l,m}^{\text{Kerr}}$ as defined by Eq.~\eqref{TensorKerrSine}.}
\label{Table_w}
\end{table}
Before analyzing the tensor, it is convenient to list some important properties associated with the integers $w_{\iota}$ that are essential for deriving its closed form representation:
\begin{enumerate}
    \item If any member of the $\{w_{\iota}\}_{\iota=1}^8$ family is odd, then the rest are also odd.
    \item Correspondingly, if any element of the $\{w_{\iota}\}_{\iota=1}^8$ set is even, then all others must also be even. This includes the $w_{\iota}=0$ case.
    \item The maximum and minimum values of $\{w_{\iota}\}_{\iota=1}^8$ are $4M$ and $1-3M$ respectively. As a result, the only possibility that gives rise to a multiple of $2M+2$ for any $\iota$ is when $\kappa_{\iota}=0,\pm 1$.
\end{enumerate}
With this at hand, one can show that the tensor now assumes the alternative and surprisingly simple form
\begin{equation}
     T^{\text{Kerr}}_{j,k,l,m}=\frac{1}{2(M+1)}\sum\limits_{\iota=1}^8(-1)^{\iota+1}\delta_{0,\rho_\iota}\,.
     \label{KerrExact2}
\end{equation}
Here $\rho_{\iota}$ denotes the remainder of the fraction $w_{\iota}/(2M+2)$ and $\delta_{0,\rho_\iota}$ is the Kronecker delta function. Equation \eqref{KerrExact2} gives an elegant analytic form of the tensor in terms of the number of supermodes $M$. From the first property of the above list, one can see that $T_{j,k,l,m}^{\text{Kerr}}$ vanishes when (i) one index from the set $\{j,k,l,m\}$ is odd while the rest are even, and (ii) one integer is even and the others are odd. Thus, the only non-zero entries of the tensor would arise only from cases where the elements of the set $\{w_{\iota}\}$ are even (zero included). To derive the non-zero elements of the tensor, we distinguish between two scenarios: (a) $w_1=2M+2$ and (b) $w_1$ being an arbitrary even integer not equal to $2M+2$. Note that these are the only relevant choices imposed by the upper and lower bounds on $w_1$ as well as the tensorial structure of $\gamma^{(\iota)}_{j,k,l,m}$ (see Eq.~\eqref{KerrExactComponents}). The alternative (a) would lead to $\gamma^{(1)}_{j,k,l,m}=2M$. This ``root", in turn, gives rise to the tree structure shown in Fig.~\ref{Tree} (left side). In particular, for each subsequent $w_{\iota}$ a tree branch is constructed via the following rules:
Determine whether $w_2$ leads to $\gamma^{(2)}=2M$ or negative $2$ which can be established using the upper and lower bounds of $w_2$ (taking into account the constraint imposed by $w_1=2M+2$). Once this is accomplished, this process is repeated. That is to say, find all possible combinations of the rest of $\{w_{\iota}\}_{\iota=3}^8$ (conforming to all constraints listed in Table~\ref{Table_w} and, in doing so, record the respective values of their auxiliary tensor $\gamma^{(\iota)}_{j,k,l,m}$. The aforementioned procedure is implemented for option (b) which gives rise to $\gamma^{(1)}=-2$ and the corresponding tree structure (see Fig.~\ref{Tree}). 
After some algebra it can be shown that the tensor values belong to the set
\begin{equation}
    S=\Big\{0,\, \frac{\pm1}{2M+2},\, \frac{1}{M+1},\, \frac{3}{2M+2}, \,\frac{2}{M+1}\Big\}\, ,
    \label{PossibleValues}
\end{equation}
where, as a reminder, $M$ denotes the number of supermodes. It is remarkable that the elements of the Kerr tensor appear in such a simple and elegant way, especially on their dependence on the number of supermodes. With these results at hand, equation \eqref{SuperModeDNLSE} (for even $M$) reads
\begin{widetext} 

    \begin{equation}
    i\frac{dc_j}{dz}+\epsilon_jc_j
    +\frac{1}{M+1}\left(\frac{3}{2}\sum_{\substack{\text{branch}\#\\2,3,5,10}}c_kc_lc_m^*+\sum_{\substack{\text{branch}\#4,6,7,\\11,12,15}}c_kc_lc_m^*
    +\frac{1}{2}\sum_{\substack{\text{branch}\#\\8,13,16,18}}c_kc_lc_m^*-\frac{1}{2}\sum_{\substack{\text{branch}\#\\9,14,17,19}}c_kc_lc_m^*\right)=0\,,
    \end{equation} 

\end{widetext}
where $\#$ labels the branch number appearing in Fig.~\ref{Tree} oriented from left to right (with branch $\#1$ appearing on the far left while branch $\#20$ to the far right). In what follows, we list a few examples of the above system. When $M=2$ we get
\begin{equation}
    i\frac{dc_1}{dz}+\underbrace{(\epsilon_1+ P-|c_1|^2/2)}_{\text{non-mixing}}c_1+\underbrace{c_2^2c_1^*/2}_{\text{mixing}}=0\, ,
    \label{M2_1}
\end{equation}
\begin{equation}
    i\frac{dc_2}{dz}+\underbrace{(\epsilon_2+ P-|c_2|^2/2)}_{\text{non-mixing}}c_2+\underbrace{c_1^2c_2^*/2}_{\text{mixing}} =0\,,
    \label{M2_2}
\end{equation}
where again $P=|c_1|^2+|c_2|^2$ is the total power (which is a conserved quantity). 
If one only keeps the non-mixing terms, we find $\langle|c_j(z)|^2\rangle=|c_j(0)|^2$ which indicates lack of thermalization. Thus, the middle terms in Eq.~\eqref{M2_1} and Eq.~\eqref{M2_2}, do not contribute to the wave mixing process. However, the presence of the last terms produces a nonlinear wave mixing, which would prohibit $\langle|c_j(z)|^2\rangle$ from remaining constant. The equations for the projection coefficients $c_j$ become slightly involved when the number of supermodes increases. This can be seen for the $M=3$ case (see Appendix B for more details). Here, one can also identify the non-mixing (${\cal N}_j$) and mixing terms (${\cal M}_j$) for which the evolution of $c_j$ is governed by
\begin{eqnarray}
    i\frac{dc_j}{dz}&+&{\cal N}_jc_j+{\cal M}_j(c_j)=0\, .
    \label{M3_j}
\end{eqnarray}
As the number of supermodes increases, the equation of motion governing the dynamics of the projection coefficients becomes more involved. This is clearly demonstrated in Fig.~\ref{TensorKerr_M5} for five supermodes.
This example clearly illustrates the important symmetry structure of the Kerr tensor mentioned earlier along with the values it assumes. Below, we summarize several properties associated with $T_{j,k,l,m}^{\text{Kerr}}$:
\begin{itemize}
    \item As one can see in Fig.~\ref{TensorKerr_M5}, at least half of the tensor elements are zero. This is due to the first property of the set $\{w_{\iota}\}$ i.e., if any member of the $w_{\iota}$ family is odd, then the rest are also odd. The locations where the tensor vanishes can be identified when an individual index from the set $\{j,k,l,m\}$ is odd (even) while the rest are even (odd). 
    \item The tensor elements are $0,\pm1/12, 1/4, 1/6, 1/3$ which coincides with the set $S$ when $M=5$. These numerical values can be found from the tree structure shown in Fig.~\ref{Tree}.
    \item For any two fixed indices, the resulting tensor slice forms a Hermitian matrix. This property is due to the high symmetry of the tensor $T_{jklm}^{\text{Kerr}}$. As we shall later see, such symmetry is absent for other cubic lattices.
    \item The tensor shown in Fig.~\ref{TensorKerr_M5} obeys the quasi-Hermiticity and permutation symmetries mentioned in Sec.~\ref{sec2_3} i.e., invariance under the transformations $k\leftrightarrow l$, $j\leftrightarrow m$, and  $k\leftrightarrow m$,  $l\leftrightarrow j$.
\end{itemize}
From the above observations, we expect the modal occupancies $|c_j|^2$ to follow a Rayleigh-Jeans distribution once thermal equilibrium is reached. In fact, we have performed numerical simulations using Eq.~\eqref{SuperModeDNLSE} for $M=20$. The results are depicted in Fig.~\ref{M20_SupermodeKerr}. This in turn confirms the theory of optical thermodynamics which also agrees with the results obtained from direct simulations using the local base \cite{MidyaFanPawel}.
\begin{figure}[!hbt]
    \centering
    \includegraphics[width=0.9\linewidth]{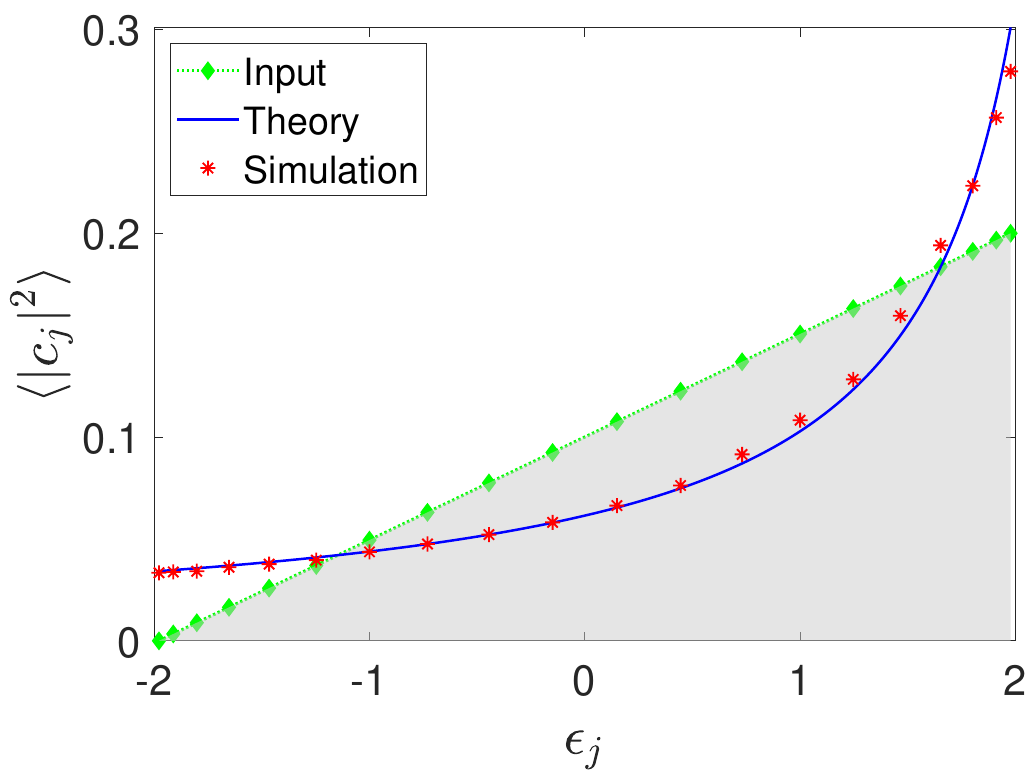}
    \caption{Modal occupancies versus eigenvalues for the Kerr lattice with power $P=2$ and energy $H_{0}=-1.9$. The initial power distribution among the various supermodes is shown in green diamonds, see Eq.~\eqref{LinearDistribution}. The numerical results (red stars) indicate the modal occupancies averaged over 400 realizations of random phases evaluated at propagation distance $z=10000$. These results were obtained by simulating equation~\eqref{SuperModeDNLSE} where the (Kerr) tensor $T_{j,k,l,m} $ is given by Eq.~\eqref{KerrExact2} (see Fig.~\ref{Tree} that helps construct the tensor values). The theory of optical thermodynamics for those values predicts a temperature $T=0.153$ and a chemical potential $\mu=-2.48$. In this case the Rayleigh–Jeans distribution Eq.~\eqref{RJD} is also shown in a solid blue line.}
    \label{M20_SupermodeKerr}
\end{figure}
To this end, we point out that the tensor analysis presented here will lay the foundation for our study of thermalization in random tensors discussed in detail in Sec.~\ref{sec6}.
\subsection{\label{sec5_2} Nonlocal lattices}
In this section we embark on the task to build an analytic form of the mixing tensor for several nonlocal optical lattices associated with Eq.~\eqref{FullLattice}. The main motivation is the identification of the tensorial symmetries which will be later tied to thermalization.
\subsubsection{\label{sec5_2_1} Case I}
\begin{figure*}[!hbt]
    \centering
    \includegraphics[width=\linewidth]{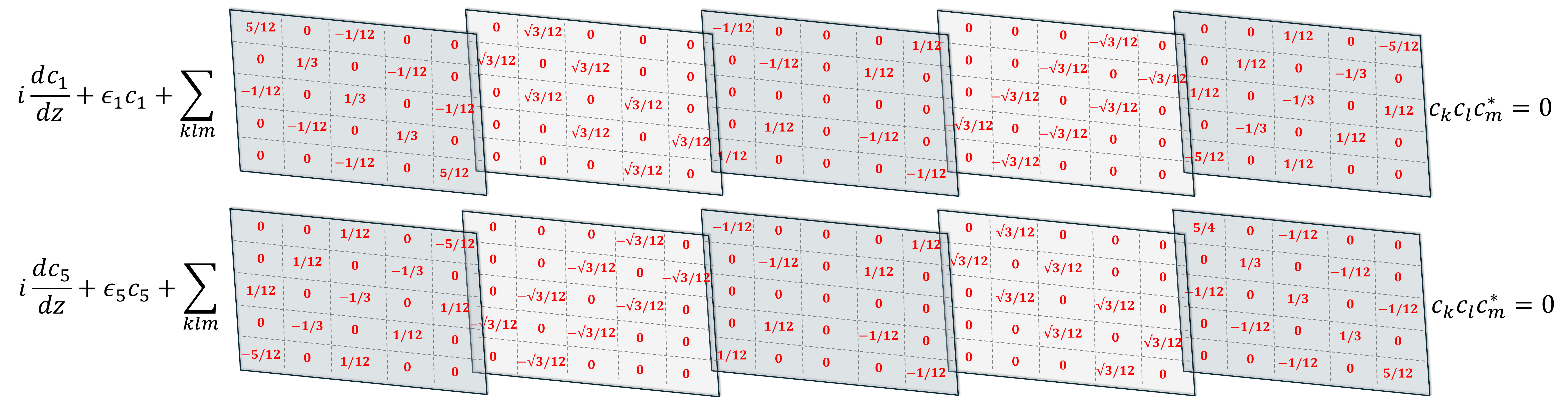}
    \caption{An illustrative example of the evolution equation~\eqref{SuperModeDNLSE} with the tensor $T_{j,k,l,m}$ given in Case I of Sec.~\ref{sec5_2} with $M=5$. For the ease of presentation we only show the cases with $j=1$ and $j=5$.}
    \label{Tensor_Non_local}
\end{figure*}

We first consider the case where $a_1=1,\,\,g=a_2=b_1=b_2=0$. In this situation, the evolution of the optical field $A_n$ is governed by
\begin{equation}
   i\frac{dA_n}{dz}+A_{n-1}+A_{n+1}+A_n\left(|A_{n-1}|^2+|A_{n+1}|^2\right)=0 \, .
   \label{NonLocal_Evolution}
\end{equation}
The nonlocal nonlinear terms appearing in Eq.~\eqref{NonLocal_Evolution} adhere to the general form presented in Eq.~\eqref{cubicTerms}, with $n_1=0, \, n_2=n_3=\pm 1$. As such, the tensor given in Eq.~\eqref{tensorExample} will have two contributions leading to
\begin{eqnarray}
    T_{j,k,l,m} &=& B 
    \sum\limits_{n=1}^{M} \left(\prod_{i=1}^2 \sin(q_i x_n) \sin(q_{i+2}x_{n+1})
    \right.
    \nonumber\\
    &+&
    \left. \prod\limits_{i=1}^2 \sin(q_i x_{n+1}) \sin(q_{i+2}x_{n})\right) \, ,
    \label{TensorNonLocalSine1}
\end{eqnarray} 
where $x_n=n\pi/(M+1)$. As a reminder, we refer to Eq.~\eqref{IndicesPiQi} for the definition of $q_i$. Interestingly enough, one can relate the tensor in Eq.~\eqref{TensorNonLocalSine1} to the Kerr one as follows:
\begin{eqnarray}
    T_{j,k,l,m} &=& 2T_{j,k,l,m}^{\text{Kerr}}\cos(x_l)\cos(x_m)\nonumber\\
    &+& 2\Gamma^{(1)}_{j,k,l,m}\sin(x_l)\sin(x_m),
\end{eqnarray}
where, 
\begin{equation}
    \Gamma^{(1)}_{j,k,l,m} =B\sum\limits_{n=1}^M \prod_{i=1}^{2}\sin\left(q_ix_n\right)\,\cos\left(q_{i+2}x_n\right)\, .
    \label{HalfCosineTensor}
\end{equation}
One can further simplify Eq.~\eqref{HalfCosineTensor} to obtain an alternative closed form similar to the Kerr case given in Eq.\eqref{KerrExact2}. Doing so, we get
\begin{equation}
     \Gamma^{(1)}_{j,k,l,m} = -\frac{B}{16}\sum_{\iota=1}^4 (-1)^{\iota+1}\left(\gamma_{j,k,l,m}^{(\iota)}-\gamma_{j,k,l,m}^{(\iota+4)}\right)\, .
     \label{HalfCosineExact}
\end{equation}
Here, the auxiliary tensor $\gamma_{j,k,l,m}^{(\iota)}$ is defined in Eq.~\eqref{KerrExactComponents}.
From Eq.~\eqref{HalfCosineExact} and the tree structure depicted in Fig.~\ref{Tree}, it can be shown that the tensor elements associated with $\Gamma^{(1)}_{j,k,l,m}$ assume the values given by the set
\begin{equation}
    {\cal S}=\left\{0, \frac{\pm1}{M+1},\frac{\pm1}{2M+2}\right\}\, .
    \label{SetForGammas}
\end{equation}
With this result at hand, an example that illustrates the structure of the tensor $T_{j,k,l,m}$ is shown in Fig.~\ref{Tensor_Non_local}. Scrutinizing Eq.~\eqref{TensorNonLocalSine1} reveals that the tensor satisfies the two symmetry conditions given in Eqs.~\eqref{QuasiHermiticity} and \eqref{Permutation}. With this fact at hand, we pose the question: Will the lattice given in Eq.~\eqref{NonLocal_Evolution} thermalize to a RJ or not? To answer this, we have numerically solved Eq.~\eqref{NonLocal_Evolution} subject to the initial random distribution given in~\eqref{LinearDistribution}. Our findings are shown in Fig.~\ref{AllRJ} (a). One can see that a RJ distribution has been reached validating our tensorial symmetry-thermalization connection.
\subsubsection{\label{sec5_3_2} Case II}
We next consider the DNLS model given in Eq.~\eqref{FullLattice} with $a_2= 1$ while setting the rest of the parameters to zero. Thus, we have
\begin{equation}
    i\frac{dA_n}{dz}+A_{n-1}+A_{n+1}+A_n^*\left(A_{n-1}^2+A_{n+1}^2\right)=0 \, .
   \label{NonLocal2_Evolution}
\end{equation}
In this case, the mixing tensor assumes the form
\begin{eqnarray}
    T_{j,k,l,m} &=& B 
    \sum\limits_{n=1}^{M}\left( \prod_{i=1}^2 \sin(p_i x_n) \sin(p_{i+2}x_{n+1})
    \right.
    \nonumber\\
    &+&
    \left. \sum\limits_{n=1}^{M} \prod_{i=1}^2 \sin(p_i x_{n+1}) \sin(p_{i+2}x_{n})
    \right) \, ,
    \label{TensorNonLocalSine2}
\end{eqnarray} 
where $\{p_{i}\}_{i=1}^4 \equiv \{ j,m,k,l\}$.
In a similar fashion to the previous nonlocal case, one can express the above lattice mixing tensor in terms of the Kerr one. That is,
\begin{eqnarray}
    T_{j,k,l,m} &=& 2T_{j,k,l,m}^{\text{Kerr}}\cos(x_k)\cos(x_l)\nonumber\\
    &+& 2\Gamma^{(2)}_{j,k,l,m}\sin(x_k)\sin(x_l),
\end{eqnarray} 
where, 
\begin{equation}
    \Gamma^{(2)}_{j,k,l,m} =B\sum\limits_{n=1}^M \prod_{i=1}^{2}\sin\left(p_ix_n\right)\,\cos\left(p_{i+2}x_n\right)\, ,
    \label{HalfCosineTensor2}
\end{equation}
which after some simplifications becomes
\begin{eqnarray}
     \Gamma^{(2)}_{j,k,l,m} &=& -\frac{B}{16}\sum_{\iota=1}^2 \left(\gamma_{j,k,l,m}^{(\iota)}+\gamma_{j,k,l,m}^{(\iota+4)}\right) \nonumber \\
     &+&\frac{B}{16}\sum_{\iota=1}^2\left(\gamma_{j,k,l,m}^{(\iota+2)}+\gamma_{j,k,l,m}^{(\iota+6)}\right)\, .
     \label{HalfCosineExact2}
\end{eqnarray}
The entries of the tensor $\Gamma^{(2)}_{j,k,l,m}$ also belong to the set ${\cal S}$ given in Eq.~\eqref{SetForGammas}. The thermalization properties and their connection to the tensor symmetries will be discussed in detail in Sec.~\ref{sec6}.

\subsubsection{\label{sec5_4_3} Case III}
Here, we consider an alternative type of nonlocal lattice governed by Eq.~\eqref{FullLattice} for which $b_1=1$ and the rest of the model parameters are set to zero. In such scenario, the evolution equation for the optical field envelope $A_n$ in local base reads
\begin{eqnarray}
    i\frac{dA_n}{dz}&+&A_{n-1}+A_{n+1}
    +A_n^2A_{n+1}^*\nonumber\\ 
    &+& 2|A_n|^2A_{n+1}+|A_{n-1}|^2A_{n-1}=0\, .
    \label{NonlocalCaseIII_localBase}
\end{eqnarray}
When viewed in supermode space, the resulting tensor appearing in Eq.~\eqref{SuperModeDNLSE} assumes the form
\begin{align}
    T_{j,k,l,m} &= B \sum\limits_{n=1}^{M}\left(
     \prod_{i=1}^3 \sin(q_i x_n) \sin(m x_{n+1})
    \right.
    \nonumber\\
    &+
    \left. \prod_{i=2}^4 \sin(q_i x_{n})\sin(j x_{n+1})\right.\nonumber\\
    &\left.+2\prod_{i=1}^3 \sin(p_i x_{n})\sin(l x_{n+1}) 
    \right) \, .
    \label{TensorNonLocalSine3}
\end{align} 
In order to identify the underlying symmetries of the tensor, we can split the last product in Eq.~\eqref{TensorNonLocalSine3} as a sum of two contributions 
that would uncover the required permutation and quasi-Hermiticity symmetries. In terms of the Kerr tensor, Eq.~\eqref{TensorNonLocalSine3} assumes the alternative form
\begin{equation}
    T_{j,k,l,m}= T_{j,k,l,m}^{\text{Kerr}}\sum_{i=1}^4\cos(x_{q_i})+\sum_{i=1}^4\Xi_{j,k,l,m}^{(i)}\sin(x_{q_i})\, ,
\end{equation}
with
\begin{equation}
    \Xi_{j,k,l,m}^{(i)}=-\frac{B}{8}\sum_{\iota = 1}^8\beta_{\iota}^{(i)}\zeta_{j,k,l,m}^{(\iota)}\, .
    \label{Xi_Tensor}
\end{equation}
Here,
$\beta^{(1)}_{\iota}=1$, when $\iota=1,4,5,8$ and $-1$ otherwise; $\beta^{(2)}_{\iota}=1$, for $\iota=1,2,7,8$ and $-1$ for the remaining cases;
$\beta^{(3)}_{\iota}=1$, if $\iota=1,3,6,8$, else it is equal to $-1$ and finally
$\beta^{(4)}_{\iota}=\pm 1$, where plus/minus sign corresponds to odd/even integer $\iota$ respectively. Furthermore, the auxiliary tensor $\zeta^{(\iota)}$ is given by
\renewcommand{\arraystretch}{1.5}
\rowcolors{1}{blue!0}{gray!0}
\begin{equation}
\zeta_{j,k,l,m}^{(\iota)} = 
\Bigg\{
    \begin{array}{lr}
        \cot\left(\frac{\pi w_{\iota}}{2(M+1)}\right), & \text{if}\,\, w_\iota\,\,\text{odd,}\\
        0, & \text{if}\,\, w_\iota\,\,\text{even,}
    \end{array}
    \label{NonLocalExactComponents}
\end{equation}
where, again, the $w_{\iota}$ are given in Table~\ref{Table_w}. A detailed derivation of the above results can be found in Appendix B. A few typical tensor slices for a small number of supermodes ($M=3$) are
\begin{equation}
    T_{1,1,l,m}=\frac{1}{4}
    \begin{pmatrix}
        3\sqrt{2} & 1/2 & -\sqrt{2}/2\\
        1/2 & \sqrt{2} & 3/2\\
        -\sqrt{2}/2 & 3/2 &0
\end{pmatrix}
,
\end{equation}
\begin{equation}
    T_{1,2,l,m}=\frac{1}{4}
    \begin{pmatrix}
        1/2 & \sqrt{2} & 3/2\\
        \sqrt{2} & 1 & 0\\
        3/2 & 0 & -3/2
\end{pmatrix}
,
\end{equation}
\begin{equation}
    T_{1,3,l,m}=\frac{1}{4}
    \begin{pmatrix}
        -\sqrt{2}/2 & 3/2 & 0\\
        3/2 & 0 & -3/2\\
        0 & -3/2 & \sqrt{2}/2
\end{pmatrix}
.
\end{equation}
It is remarkable that the entries of these tensor slices are rather simple looking which is unexpected given the fact that the type on nonlinearity in local space is rather non-trivial. This pattern also shows up if one slices the tensor along a different hyperplane. 
\subsubsection{\label{sec5_5_4} Case IV}
Lastly, we analyze the structure of the nonlinear mixing tensor for the generalized DNLS equation~\eqref{FullLattice}. Here we take $b_2=1$ and choose the rest of the model parameters to zero. Under such assumptions, the governing dynamical system is reduced to 
\begin{eqnarray}
    i\frac{dA_n}{dz}&+&A_{n-1}+A_{n-1}
    +A_n^2A_{n-1}^*\nonumber\\ 
    &+& 2|A_n|^2A_{n-1}+|A_{n+1}|^2A_{n+1}=0\, .
    \label{CaseIV_evolution}
\end{eqnarray}
Following similar analysis discussed in previous cases, after some algebra we arrive at 
\begin{eqnarray}
    T_{j,k,l,m} &=& B 
    \sum\limits_{n=1}^{M} \left(\prod_{i=1}^3 \sin(q_i x_n)\sin(m x_{n-1}) 
    \right.
    \nonumber\\
    &+&
    \left. \prod_{i=2}^4 \sin(q_i x_{n})\sin(j x_{n-1})
    \right.
    \nonumber\\
    &+&
    \left.2\prod_{i=1}^3 \sin(p_i x_{n}) \sin(l x_{n-1})
    \right) \, .
    \label{TensorNonLocalSine4}
\end{eqnarray} 
Using similar arguments as discussed in Case III one can establish the validity of the quasi-Hermiticity and permutation symmetries of the tensor $T_{j,k,l,m}$. It is again interesting to note that the tensor under discussion is related to its Kerr counterpart. As such, we have
\begin{equation}
    T_{j,k,l,m}= T_{j,k,l,m}^{\text{Kerr}}\sum_{i=1}^4\cos(x_{q_i})-\sum_{i=1}^4\Xi_{j,k,l,m}^{(i)}\sin(x_{q_i})\, .
    \label{TensorNonlocal4WithKerr}
\end{equation}
It is evident that the tensor structure in Eq.~\eqref{TensorNonlocal4WithKerr} is intricate which makes its simplification a rather formidable task. Nonetheless some representative examples of tensor slices when $M=3$ that highlight its intrinsic structure are given:
\begin{equation}
    T_{1,1,l,m}=\frac{1}{4}
    \begin{pmatrix}
        3\sqrt{2} & -1/2 & -\sqrt{2}/2\\
        -1/2 & \sqrt{2} & -3/2\\
        -\sqrt{2}/2 & -3/2 &0
\end{pmatrix}
,
\end{equation}
\begin{equation}
    T_{1,2,l,m}=\frac{1}{4}
    \begin{pmatrix}
        -1/2 & \sqrt{2} & -3/2\\
        \sqrt{2} & -1 &  0\\
        -3/2 & 0 & 3/2
\end{pmatrix}
,
\end{equation}
\begin{equation}
    T_{1,3,l,m}=\frac{1}{4}
    \begin{pmatrix}
        -\sqrt{2}/2 & -3/2 & 0\\
        -3/2 & 0 & 3/2\\
        0 & 3/2 & \sqrt{2}/2
\end{pmatrix}
.
\end{equation}
Similar arguments can be made about the simplicity of these tensor slices given the fact that a priori one would have expected more complicated results. 

In the next section, we will address the issue of ``correlation" between the above-mentioned symmetries and equilibration to a RJ distribution of all the nonlocal models discussed above.
\section{\label{sec6}Thermalization of cubic nonlinear Lattices}

\subsection{\label{sec6_1}Connections to tensorial symmetries}
One of the main goals of the current study is to investigate a possible connection between the symmetries of the tensors $T_{j,k,l,m}$ that appear in Eq.~\eqref{SuperModeDNLSE} and the thermalization properties of the corresponding cubic nonlinear lattices. In other words, we would like to pose the question whether nonlinear tensors preserving the quasi-Hermiticity and permutation symmetries assumed in Eq.~\eqref{QuasiHermiticity} and \eqref{Permutation} lead to a Rayleigh-Jeans equilibrium distribution. If this unidirectional hypothesis (tensorial symmetries imply thermalization) holds, it would provide a useful tool to investigate thermalization of cubic lattices solely by scrutinizing their nonlinear tensorial symmetries.
\begin{figure}[!hbt]
    \centering
    \includegraphics[width=\linewidth]{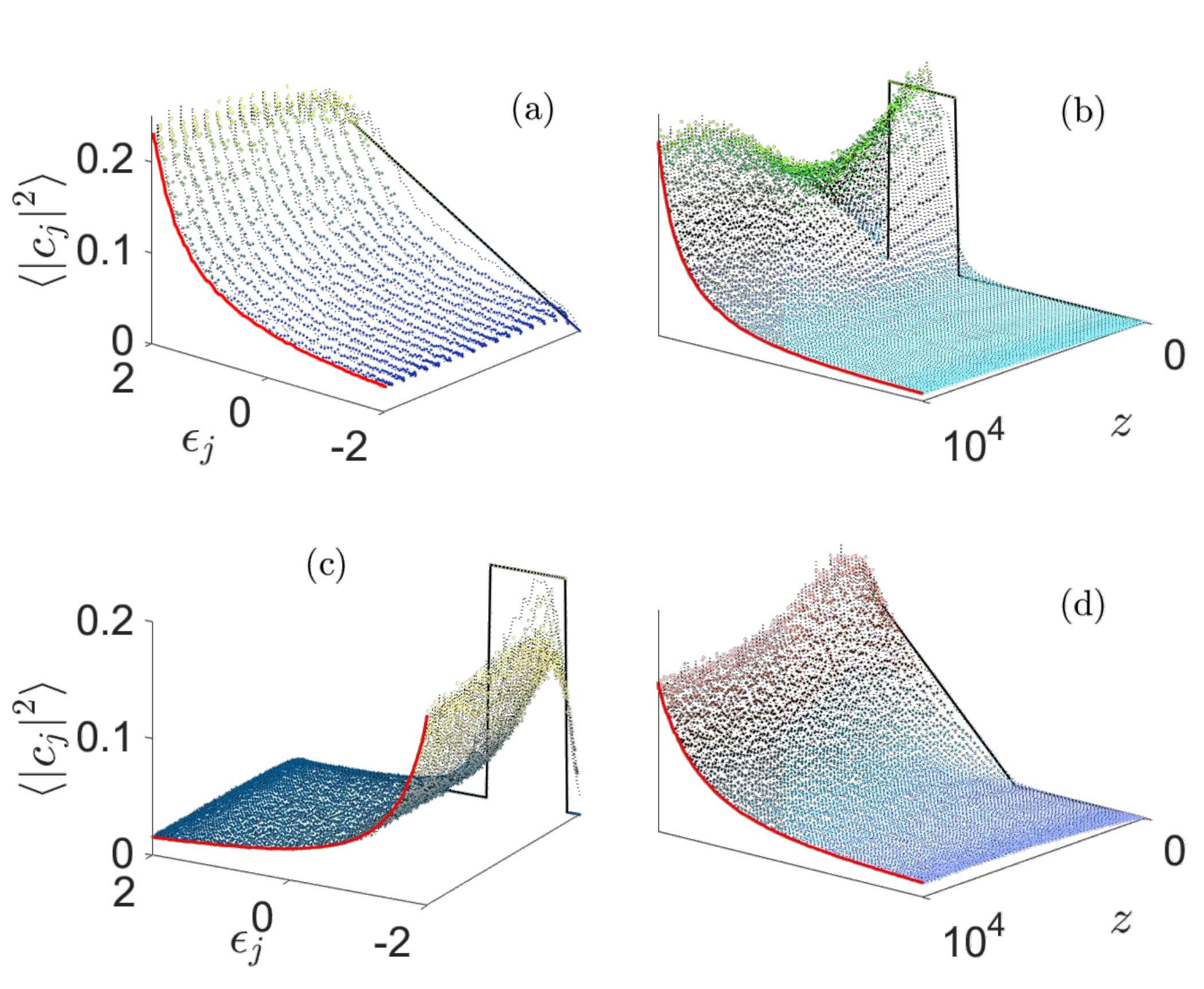}
    \caption{Evolution of the ensemble averaged modal occupancies for various nonlocal nonlinear lattices with $M=100$ supermodes. (a) The first nonlocal lattice (Case I) with total power of $P=8$ and linear initial distribution across the modes (indicated by a solid black line at $z=0$). The lattice approaches a Rayleigh-Jeans distribution (red line at $z=10000$), agreeing with the theoretically predicted values for the temperature $T=0.19$ and a chemical potential $\mu=-2.7$. (b) and (c) correspond to the nonlocal cases II and III respectively. Here, the initial condition corresponds to an equally distributed power of $P=8$ among the supermodes $65\leq j \leq 89$ (b) and $P=5$ among $15\leq j \leq 39$ (c) supermodes. Both simulations lead to a RJ distribution with temperatures and chemical potentials $T=0.066$,  $\mu=-2.15$ (b) and $T=-0.1$, $\mu= -0.7$(c) as the theory predicts. Lastly part (d) corresponds to the nonlocal case IV with power $P=5$ linearly distributed (at $z=0$) across the supermodes with indices $50\leq j \leq 100$. The temperature at thermal equilibrium is $T=0.07$ and the chemical potential $\mu=-2.4$. The red solid line indicates the RJ distribution at $z=10^4$.}
    \label{AllRJ}
\end{figure}
With this in mind, we shall use the discrete model Eq.~\eqref{FullLattice} as a testbed example for which all the underlying tensorial structures have been analytically derived. Furthermore, these tensorial symmetry conditions can be readily verified by inspecting their exact closed from provided in Sec.~\ref{sec5}. To this goal, we next perform direct numerical simulations on Eq.~\eqref{FullLattice} with $V_n=0$ corresponding to random initial conditions of the form: 
\begin{equation}
    A_n(0)=\sum\limits_{j=1}^M \tilde c_je^{2\pi i r_j}\psi_n^{(j)}\, ,
\end{equation}
where $r_j$ is a random field uniformly distributed on the interval $(0,1)$ and $\tilde c_j$ are deterministic modal amplitudes defined by either Eq.~\eqref{LinearDistribution} or~\eqref{EquipartitionDistribution}.
Bellow, we report on our numerical findings for all lattices (Kerr and nonlocal) in the same order as presented in Sec.~\ref{sec5}.
\begin{enumerate}
    \item \underline{Kerr Lattice}.  The Hamiltonian is given by
    \begin{equation}
        H =H_0+\frac{1}{2}\sum\limits_{n=1}^M|A_n|^4\, ,
    \end{equation}
    with $H_0$ being the part corresponding to the nearest neighbor coupling. Thermalization properties of the Kerr lattice have been thoroughly studied over the past several years \cite{MidyaFanPawel,Kottos2011}. It is well known that equilibration to a RJ distribution has been reported (see for example Fig.~\ref{M20_SupermodeKerr}). The Kerr tensor given in Eq.~\eqref{TensorKerrSine} does satisfy the quasi-Hermiticity and permutation symmetries postulated. As such, the Kerr nonlinear lattice conforms with this hypothesis. It is interesting to note that the Kerr tensor exhibits high symmetry, i.e., invariance under any index permutation. Compared with the Kerr case, all other nonlocal tensors admit reduced symmetries.   
    \item \underline{Nonlocal Lattices}. 
\end{enumerate}
\begin{itemize}
    \item \underline{Case I}. This is the first example, which we use to test a possible link between the constructed tensorial symmetries and their implication on relaxation to a RJ distribution. To do so, we consider the dynamical lattice given in Eq.~\eqref{NonLocal_Evolution} for which the corresponding Hamiltonian reads 
    \begin{equation}
        H =H_0+\sum\limits_{n=1}^M|A_n|^2|A_{n+1}|^2\, .
    \end{equation}
    Note that the resulting mixing tensor and its associated symmetries have been exactly identified in Sec.~\ref{sec5_2}. Thus, to verify the symmetry-thermalization connection, we need to resort to numerical simulations performed on the given Hamiltonian.
    The results of such computation are summarized in Fig.~\ref{AllRJ} (a), where the dependence of the quantity $\langle|c_j|^2\rangle$ on the system's linear eigenenergies is shown. As one can see, its form coincides with theoretically predicted RJ distribution which supports the above-mentioned conjecture.
    \item \underline{Case II}. The corresponding Hamiltonian for the system given in~\eqref{NonLocal2_Evolution} is given by
    \begin{equation}
         H =H_0+\frac{1}{2}\sum\limits_{n=1}^M\big(A_n^2(A_{n+1}^*)^2 + \text{c.c.}\big) \, .
    \end{equation}
    We have performed direct numerical simulations using the above Hamiltonian to check if thermalization to RJ is possible. In Fig.~\ref{AllRJ} (b) we show the ensemble averaged modal occupancies in terms of the energy eigenstates. It is clear that the system indeed relaxes to a RJ distribution. Given the fact that the tensor corresponding to this Hamiltonian has been computed in Sec.~\ref{sec5_2} (which obeys the quasi-Hermiticity and permutation symmetries), then one can positively affirm the unidirectional connection between tensorial symmetries and thermalization.
    \item \underline{Case III}. In this situation, the evolution equation of the optical field $A_n$ is governed by Eq.~\eqref{NonlocalCaseIII_localBase} which can be derived from the Hamiltonian functional
    \begin{equation}
         H = H_0+ \sum\limits_{n=1}^M   \left(A_n^*A_{n+1}^*A_n^2+\text{c.c.}\right)  \, ,
    \end{equation}
    using the standard Poisson brackets defined in Eq.~\eqref{PoissonBrac}.  
    The current lattice exhibits stronger nonlinear nonlocality than the previous two examples. Consequently, if the input power is scaled proportionally to ensure the system's Hamiltonian remains dominated by its linear component, one would expect thermalization to occur ``faster". We have studied the possibility of thermalization to a distribution that follows the Rayleigh-Jeans law by simulating the above Hamiltonian system subject to uniform modal distribution (see Fig.~\ref{AllRJ} (c)). Indeed, the system reached a RJ distribution which coincides with the theoretically predicted formula of optical thermodynamics given in Eq.~\eqref{RJD}. Interestingly enough, the corresponding tensorial symmetries found in Sec.~\ref{sec5_2} appear to conform with our symmetry-thermalization connection.
    \item \underline{Case IV}: Lastly, we probe the tensorial symmetries and their connection to thermalization by considering another type of nonlocal nonlinear lattice as given in Eq.~\eqref{CaseIV_evolution} with Hamiltonian structure
    \begin{equation}
         H = H_0+ \sum\limits_{n=1}^M   \left(A_n^*A_{n+1}^*A_{n+1}^2+\text{c.c.} \right) \, .
    \end{equation}
    As before, we performed numerical simulations on the above Hamiltonian and found that the statistically averaged modal occupancies agree with the theoretically predicted RJ distribution, see Fig.~\ref{AllRJ} (d). Furthermore, the associated tensor preserves the two postulated symmetries given in Eq.~\eqref{QuasiHermiticity} and~\eqref{Permutation}. This case provides another indication that supports the link between tensorial symmetries and RJ distribution.
\end{itemize}

\subsection{\label{sec6_2} Random tensors}
Most of the analytical and computational studies concerning thermalization of nonlinear lattices focus on dynamics defined in local space. This approach seems to be natural since the underlying physical models are always formulated in local base. In this regard, the dynamics in the modal base is less explored. One of the main reasons being the intricate structure of the tensor $T_{j,k,l,m}$ and the combinatorial sum that appears in Eq.~\eqref{SuperModeDNLSE}. For example, for a system with $M$ supermodes, the number of nonlinear terms is of the order of $M^3$ and the tensor is of size $M^4$. Nonetheless, Eq.~\eqref{SuperModeDNLSE} offers several advantages: 
\begin{itemize}
    \item Provides an elegant and unified formulation for all cubic nonlinear lattices. In other words,
    Eq.~\eqref{generic-DNLS} with an arbitrary number of nonlinear cubic terms would look the same as in Eq.~\eqref{SuperModeDNLSE} with the proper tensor.
    \item The nonlinear evolution equation in supermode base (Eq.~\eqref{SuperModeDNLSE}) allows one to probe new nonlinear systems without any direct reference to the dynamics on a local base. 
    \item Opens the opportunity to study thermalization for random tensors. This highlights the importance of the quasi-Hermiticity and permutation symmetries (which are not visible in the local base) in the theory of thermalization of nonlinear lattices. In other words, it provides a unique testbed for our hypothesis, which asserts that tensorial symmetries lead to thermalization.
\end{itemize}
The study of thermalization with random tensors provides a unique opportunity to test the hypothesis formulated earlier regarding the role that tensor symmetries play in thermalization. We thus, consider Eq.~\eqref{SuperModeDNLSE} where now the tensor is replaced by a random tensor i.e.,
\begin{equation}
    T_{j,k,l,m}\longrightarrow T^{\text{rand}}_{j,k,l,m}\, .
    \label{Random_Tensor}
\end{equation}
The elements of the tensor $T^{\text{rand}}_{j,k,l,m}$ are generated 
from a uniform probability distribution on the interval $(-0.1,0.1)$ such that they adhere to the presumed symmetry conditions: 
\begin{align}
    &\text{quasi-Hermiticity: }\,\,\,T^{\text{rand}}_{j,k,l,m} = T^{\text{rand}}_{l,m,j,k}\, ,\\
    \nonumber\\
   &\text{permutation symmetry: }\,\,\, T^{\text{rand}}_{j,k,l,m} = T^{\text{rand}}_{m,l,k,j}\, .
\end{align}
\begin{figure}[!hbt]
    \centering
    \includegraphics[width=\linewidth]{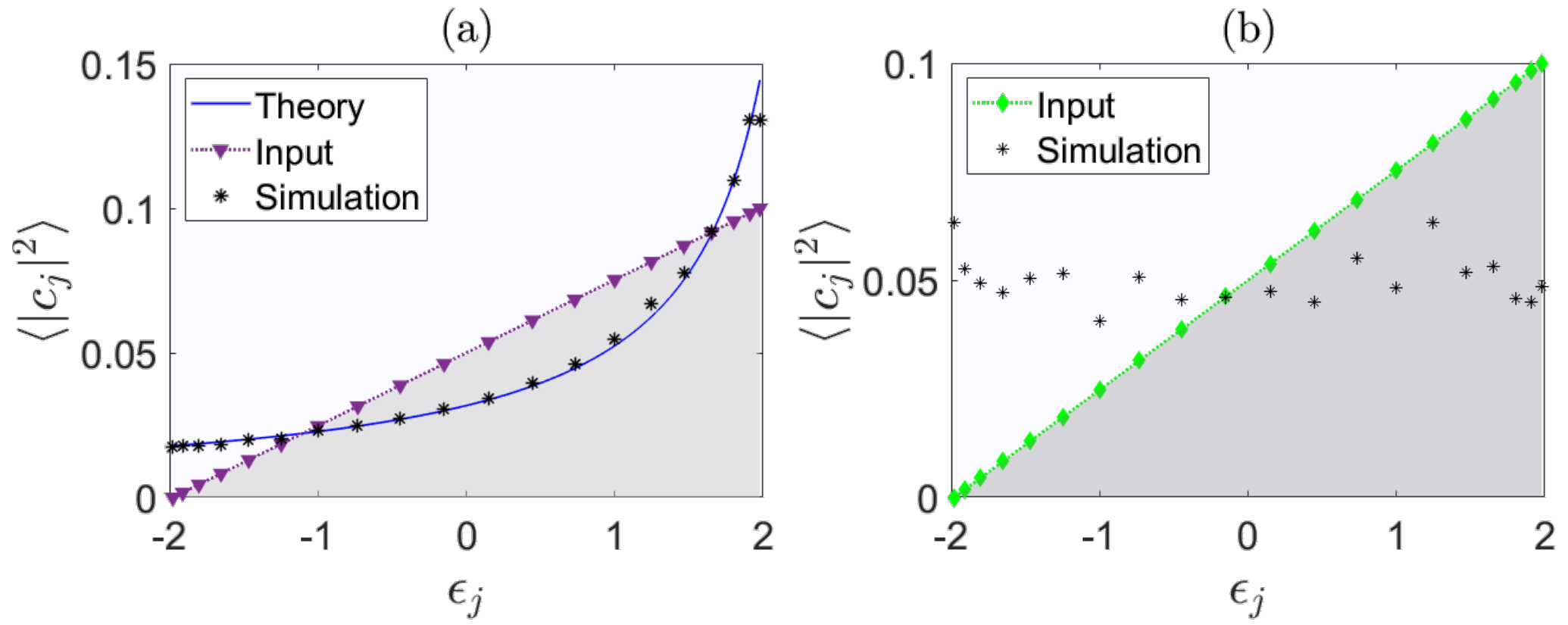}
    \caption{Modal occupancies over eigenvalues diagram for a random tensor that preserves (a) quasi-Hermiticity and permutation symmetries (b) only the quasi-Hermiticity condition. The total power $P=1$ is initially linearly distributed across $M=20$ modes in an ascending order from higher to lower order modes (purple/green line respectively). After 400 ensembles over random phase initial conditions the distribution of the modal occupancies is depicted (black stars) at $z=5000$. In (a) the solid blue line is the theoretically predicted RJ distribution with temperature $T=0.08$ and chemical potential $\mu=-2.5$. In (b) a RJ distribution is not observed. Instead, the system tends to converge toward an equilibrium state characterized by power equipartition.}
    \label{Supermode_Two_Symmetries}
\end{figure}
We have performed numerical simulations using Eq.~\eqref{SuperModeDNLSE} with the random tensor given in Eq.~\eqref{Random_Tensor}. A summary of our results is depicted in Fig. \ref{Supermode_Two_Symmetries} (a), where the ensemble averaged modal occupancies $\langle|c_j|^2\rangle$ over many realizations of initial random modal phases is shown to relax to a RJ distribution. These numerical findings highlight the important role that the tensorial quasi-Hermiticity and permutation symmetries play in thermalization of cubic lattices. It is interesting to mention that if one relaxes the permutation symmetry condition (while preserving the quasi-Hermiticity) then the system no longer approaches a RJ distribution. Instead, it reaches an equipartition state as is seen in Fig. \ref{Supermode_Two_Symmetries} (b). The results corroborate our proposition that the intrinsic tensorial symmetries play a key role in determining the ultimate functional form of the statistical distribution of the modal amplitudes, as well as in the overall process of thermalization of nonlinear cubic optical lattices.
\begin{figure*}[ht]
    \centering
    \includegraphics[width=\linewidth]{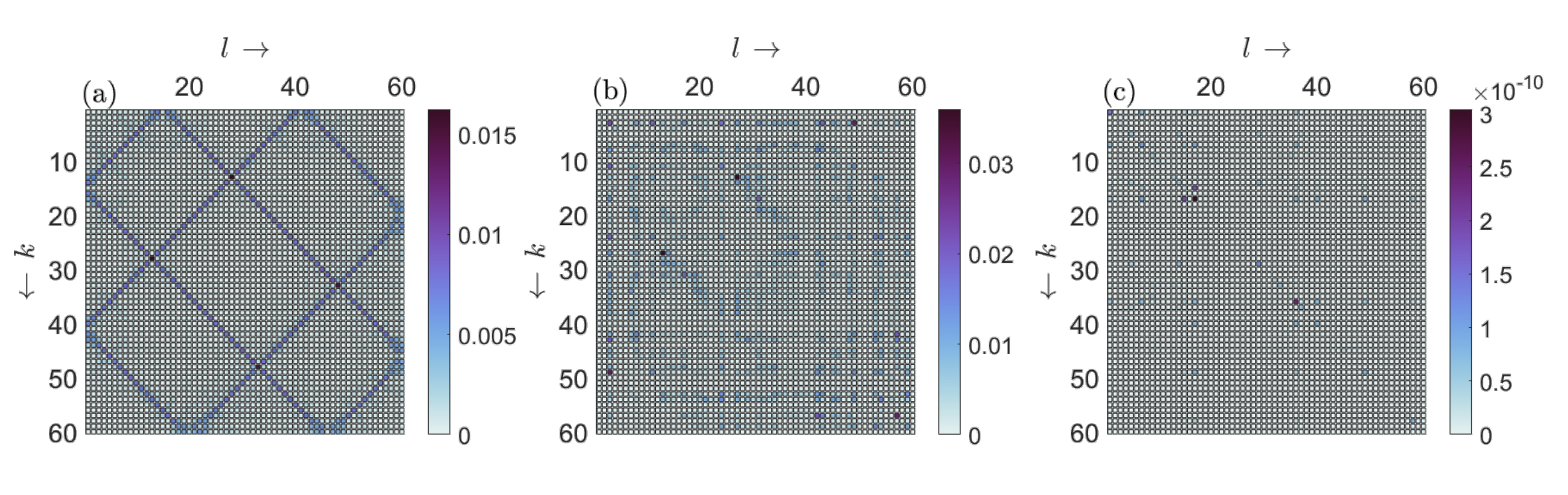}
    \caption{Typical tensor slices (in absolute value) for the nonlinear Anderson model with random potential $V_n$ uniformly distributed on the interval $(-W,W)$. (a) $W=0.1$, (b) $W=1$ (c) $W=5$. The number of supermodes is 60. The number of nonzero tensor elements decreases as the potential disorder increases resulting in a small number of nonlinear mixing terms (i.e., short range nonlinearity).}
    \label{AndersonAllSlices}
\end{figure*}
\subsection{\label{sec6_3}Dependence on supermodes}
So far, we have studied the structure of the mixing tensor $T_{j,k,l,m}$ that appears in Eq.~\eqref{SuperModeDNLSE} for various types of nonlinear lattices with nearest neighbors coupling in the absence of any external potentials. In this case, each supermode given by Eq.\eqref{RegLatEigvec} occupies all lattice sites. This leads to the important conclusion
\begin{equation*}
    \text{extended supermodes } \longrightarrow \text{extended mixing tensor}
\end{equation*}
The above statement is valid irrespective of whether the nonlinearity (in local base) is short (e.g. Kerr) or long ranged (such as the nonlinearity in Eq.~\eqref{FullLattice}). When the tensor is dense (like the examples presented in this paper), the dimension of the mixing terms in supermode base is large, causing the system to reach a RJ distribution relatively fast (see the numerical results presented in Sec.~\ref{sec6_1}). The situation becomes more intricate when the supermodes are in a localized state, as is the case when the potential $V_n$ corresponds to either the Anderson disordered model \cite{Anderson} or the Aubry-André quasi-periodic case \cite{aubryAndre}. In other words, the sparseness of the tensor depends on the localization length of the supermodes as well as on the type of nonlinearity. That is to say 
\begin{equation*}
    \text{strongly localized tensors } \longrightarrow \text{slow thermalization}\, 
\end{equation*}
In Fig.~\ref{AndersonAllSlices} we show a typical example of the mixing tensor for various strength of disorder lattice potentials uniformly distributed on the interval $(-W,W)$. As one can see, for small values of disorder ($W=0.1$), the tensor slice is in an extended state, whereas it becomes progressively more localized as the disorder strength gets larger. The punchline is the following: The less (more) the supermodes are localized, the sparser (denser) the mixing tensor becomes, which in turn forces the thermalization process to evolve slower (faster). This behavior has been observed on the Anderson model with Kerr nonlinearity \cite{PyrialakosSlowdown}.
In the next section, we shall elaborate more on this issue through the perspective of the mixing tensor.
\subsection{\label{sec6_4}Optical lattices in modal space: an inverse approach}
The standard approach to the study of thermalization of nonlinear lattices follows a prescribed dynamics in local base which often times is governed by a ``nice" set of equations, such as the ones considered in this paper. This implies that the evolution of the projection coefficients $c_j(z)$ obeys a rather complicated coupled dynamical system. As such, the number of nonlinear terms is of the order of $M^3$, making their analysis difficult. This is the case for our current study: the nonlinear interactions in local base appear in an elegant form while ``looking messy" in supermode base. Since the role of weak nonlinearity is to steer the system into an equilibrium state (whenever it exists) and the nonlinearity type is ``irrelevant" (Kerr or nonlocal) this raises the question whether one can devise a method whereby the nonlinear interactions in supermode base look ``simple" at the expense of dealing with ``unpleasant" nonlinearities in local base. Since the quantity of interest is given by $\langle|c_j|^2\rangle$, this inverse approach would provide an advantage to probe the role of mixing tensors in reaching the equilibrium state.
\begin{figure}[!hbt]
    \centering
    \includegraphics[width=\linewidth]{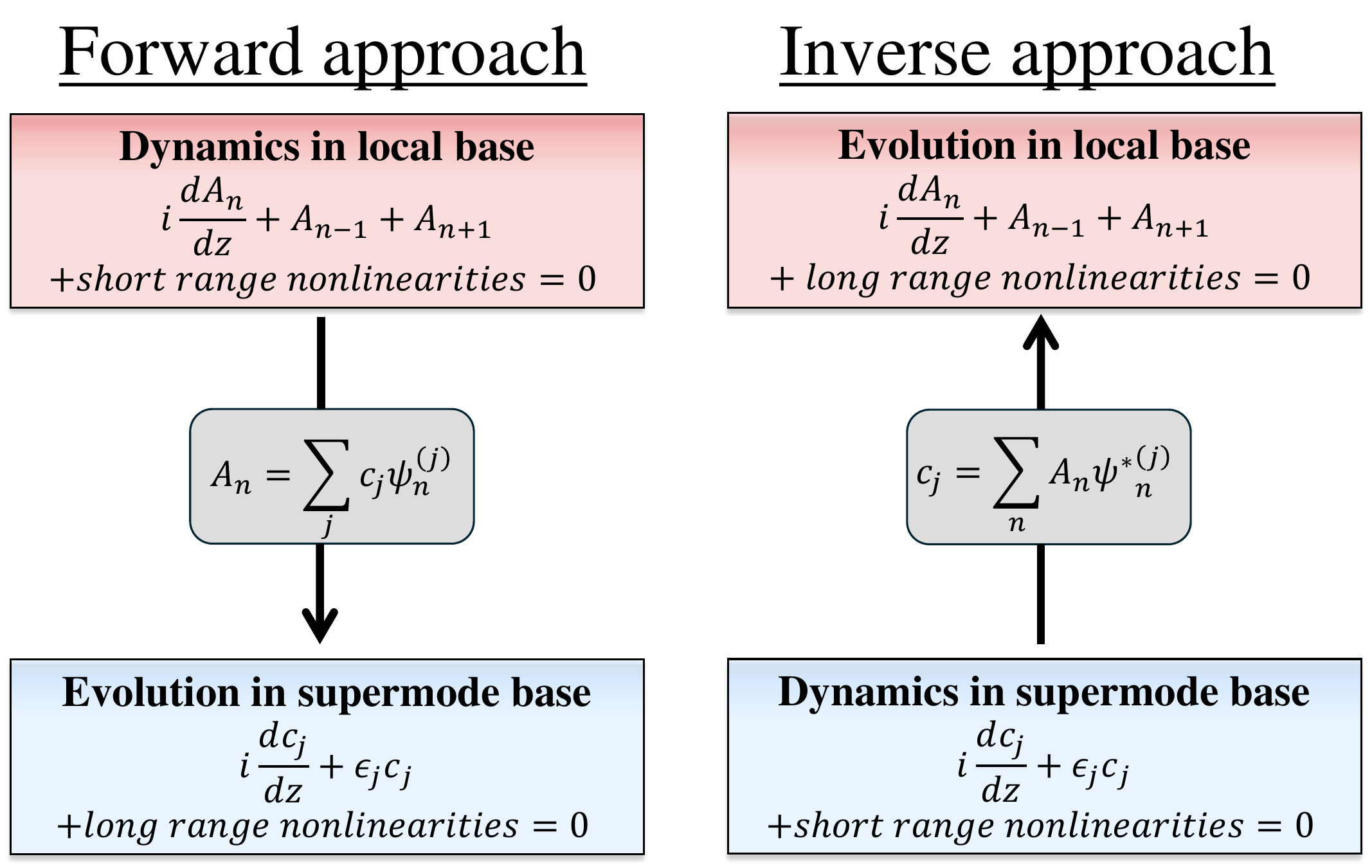}
    \caption{A schematic presentation of the forward and inverse approaches to the study of thermalization. In the conventional case, the dynamics of the wave amplitude at site $n$ in local base contains short range nonlinearities (e.g., Kerr, and/or nonlocal terms, see Eq.~\eqref{FullLattice}). Since the supermodes are in an extended state, under the transformation $A_n\rightarrow c_j$ the dynamics in supermode base contains long range nonlinear couplings. Contrary to this, if one begins with an evolution equation for the projection coefficients $c_j$ under the assumption of short range interactions then the inverse transform $c_j\rightarrow A_n$ produces long range nonlinear coupling in the local base.}
    \label{forwardinversediagram}
\end{figure}
The idea is schematically shown in Fig.~\ref{forwardinversediagram}. To this end, we aim at deriving a large class of lattice models that are embedded in the supermode base containing short range nonlinearities that preserve the power and Hamiltonian. The starting point is the evolution equation \eqref{SuperModeDNLSE} which for simplicity we write again
\begin{equation}
    i\frac{dc_j}{dz}+\epsilon_j c_j+\sum_{k,l,m=1}^{M} \mathcal{T}_{j,k,l,m}\,c_k c_l c^*_m=0\, ,
    \label{SuperModeDNLSE_again}
\end{equation}
with the lattice on-site energies $\epsilon_j$ given by either Eq.~\eqref{RegLatEigval}, Anderson random type, or the Aubry-André model. Furthermore, it is assumed that the tensor $\mathcal{T}_{j,k,l,m}$ is constructed from 
\begin{equation}
    \mathcal{T}_{j,k,l,m}=\sum_{n=1}^M\prod_{q_i=1}^4 \Phi_n^{(q_i)}\, , \, \, \{q_i\}_{i=1}^4 = \{ j,k,l,m \}\, ,
    \label{Tensor_Delta}
\end{equation}
with
\begin{equation}
    \Phi_n^{(q_i)} =\sum_{\xi=0}^{N-1}\alpha_{\xi}\delta_{n-\xi,q_i}\, ,
    \label{DeltaSupermodes}
\end{equation}
where $N\leq M$ counts the number of peaks of the wavefunction $ \Phi_n^{(q_i)}$ with real amplitude $\alpha_{\xi}$ and $\delta_{i,j}$ is the Kronecker delta function. Substituting Eqs.~\eqref{Tensor_Delta}, \eqref{DeltaSupermodes} into Eq.~\eqref{SuperModeDNLSE_again} we arrive at the evolution equation
\begin{eqnarray}
    &&i\frac{dc_j}{dz}+\epsilon_jc_j\nonumber\\
    &+&\sum_{\zeta=0}^{N-1}\left(\alpha_{\zeta}\Bigg|\sum_{\xi=0}^{N-1}\alpha_{\xi}c_{j+\zeta-\xi}\Bigg|^2\left(\sum_{\xi=0}^{N-1}\alpha_{\xi}c_{j+\zeta-\xi}\right)\right)=0\, .\nonumber\\
    \label{SupermodeDeltaDynamics}
\end{eqnarray}
Equation \eqref{SupermodeDeltaDynamics} is a novel set of dynamical systems that provides an opportunity to probe the effect of nonlinear wave mixing on thermalization processes. To reconstruct the corresponding lattice equation in local base we use the inverse approach described in Fig.\ref{forwardinversediagram}. This leads to 
\begin{equation}
    i\frac{dA_n}{dz}+A_{n+1}+A_{n-1}+\sum_{k,l,m=1}^MT_{\text{dual}}^{n,k,l,m}A_kA_lA_m^*=0\, ,
    \label{DualLocalEquatio}
\end{equation}
where
\begin{widetext}
    \begin{equation}
    T_{\text{dual}}^{n,k,l,m}=\sum_{j=1}^M\sum_{\{\xi\}=0}^{N-1}\alpha_{\xi_0}\alpha_{\xi_1}\alpha_{\xi_2}\alpha_{\xi_3}\psi_n^{(j)}\psi_k^{{(j+\xi_0-\xi_1)}^*}\psi_l^{{(j+\xi_0-\xi_2)}^*}\psi_m^{(j+\xi_0-\xi_3)}\, ,
    \label{DualTensor}
\end{equation}
\end{widetext}
and $\{\xi\}$ stands for a short notation for $\{\xi_0,\xi_1,\xi_2,\xi_3\}$. In deriving equations \eqref{DualLocalEquatio} and \eqref{DualTensor} we used $\epsilon_j$ that corresponds to the free lattice in which $\psi_n^{(j)}$ is given by Eq.~\eqref{RegLatEigvec}. Instead, one can obtain the corresponding local equation in the presence of any $\epsilon_j$ set of eigenvalues, such as the Anderson model. 
To this end, we next present several examples in support of the hypothesis that long range nonlinear couplings in supermode base facilitate nonlinear wave mixing, which eventually leads to faster thermalization.
\begin{figure}[!hbt]
    \centering
    \includegraphics[width=\linewidth]{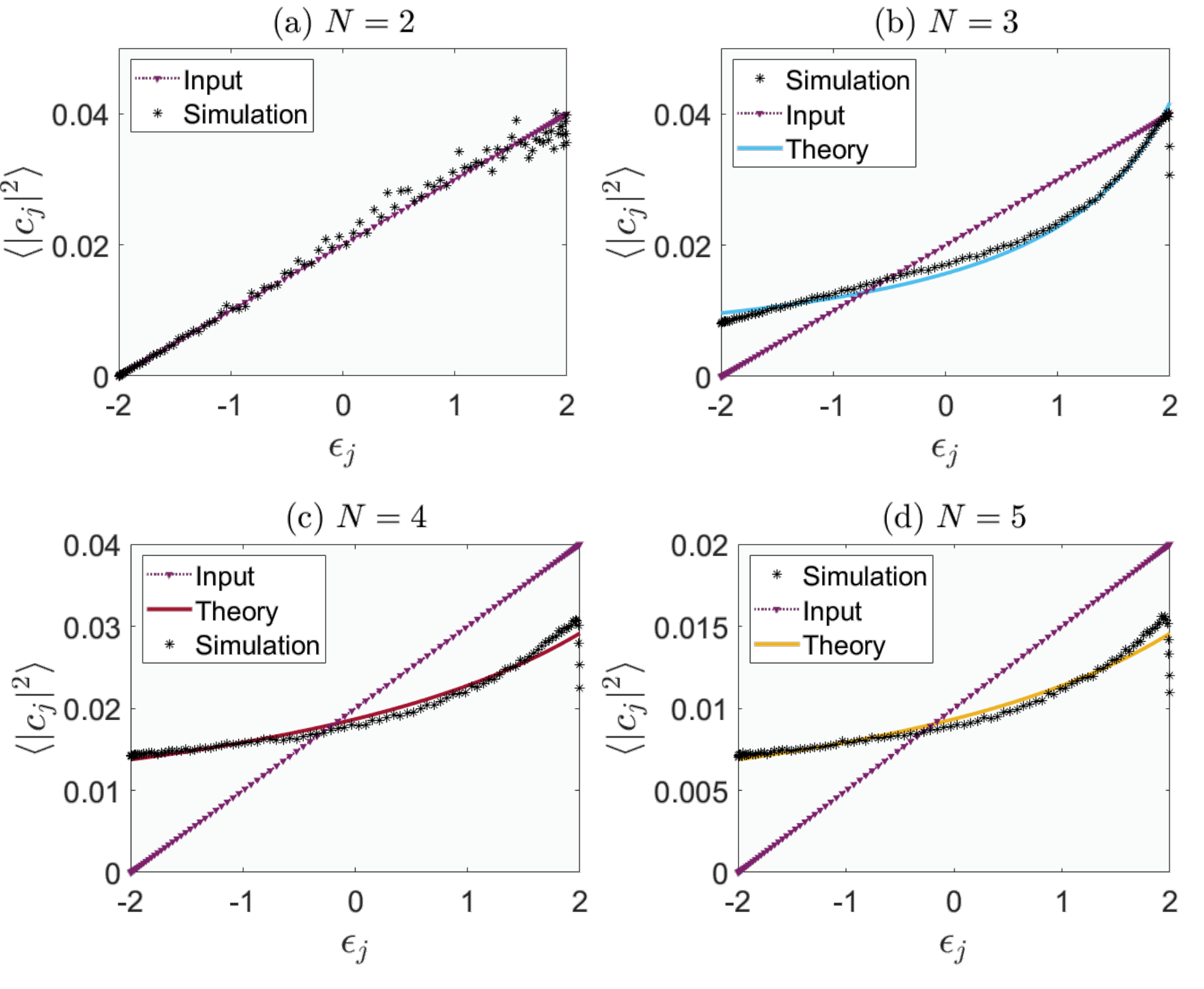}
    \caption{Numerical simulation of Eq.~\eqref{SupermodeDeltaDynamics} for various values of $N$ with a linear initial distribution among $M=100$ supermodes and eigenvalues $\epsilon_j$ given by Eq.~\eqref{RegLatEigval}. The simulation results for the modal occupancies are shown at $z=120000$ and are averaged over 800 ensembles of random phase initial conditions. (a) For $N=2$ and $P=2$ the averaged modal occupancies remain nearly unchanged from their initial distribution. (b) For $N=3$ and $P=2$ an almost exact Rayleigh–Jeans distribution (blue solid line) is attained, although some discrepancies appear for higher-order modes in the range $-2<\epsilon_j<0$. (c) For $N=4$ and $P=2$, we also get a good match between theory (red solid line) and simulation (black stars) and (d) for $N=5$ and $P=1$ the modal occupancies (black stars) match the theoretically predicted RJ distribution (solid yellow line) nearly perfectly.}
    \label{All_Delta}
\end{figure}
\begin{itemize}
    \item \underline{$N=1$}: This case corresponds to $\Phi_n^{(q_i)}$ with a single peak (extreme localization regime) for $\alpha_0=1$. As a result Eq.~\eqref{SupermodeDeltaDynamics} becomes
    \begin{equation}
        i\frac{dc_j}{dz}+\epsilon_jc_j+|c_j|^2c_j=0\, .
        \label{KerrExact}
    \end{equation}
Consequently, $|c_j|^2$ is a constant of motion and thus the initial distribution among the $M$ supermodes remains invariant, i.e., 
\begin{equation}            
    \langle|c_j(z)|^2\rangle= \langle|c_j(0)|^2\rangle\, .
\end{equation}
In other words, in the extreme localization regime there is no relaxation to a RJ distribution.
\item $\underline{N=2}$: Here, the wave functions $\Phi_n^{(q_i)}$ are composed of two peaks (for sake of simplicity are taken to be of equal heights with $\alpha_0=\alpha_1=1$). As a result, Eq.~\eqref{SupermodeDeltaDynamics} takes the surprisingly simple form: 
\begin{eqnarray}
    i\frac{dc_j}{dz}+\epsilon_jc_j&+&|c_{j}+c_{j-1}|^2(c_{j}+c_{j-1})\nonumber\\
    &+&|c_{j}+c_{j+1}|^2(c_{j}+c_{j+1})=0\, .
    \label{Supermode2Delta}
\end{eqnarray}
Unlike the previous scenario ($N=1$), here the nonlinear mixing terms are short range nonlocal, i.e., the nonlinear coupling is between field amplitudes located at sites $j$ and $j\pm1$. An important and immediate question that arises is whether Eq.~\eqref{Supermode2Delta} thermalizes to a RJ distribution or not. To answer this, we simulated Eq.~\eqref{Supermode2Delta} using a linear initial distribution among $M=100$ supermodes with $\epsilon_j$ given by Eq.~\eqref{RegLatEigval}. The results can be summarized in Fig.~\ref{All_Delta} (a).
Clearly, a highly localized supermode with two peaks leads to a highly localized mixing tensor, which prevents the system from thermalizing.
\item $\underline{N=3}$: This is an interesting case where one starts to observe a change in the modal statistical distribution. Similar to the previous analysis one arrives at the coupled system:
\begin{eqnarray}
    i\frac{dc_j}{dz}&+&\epsilon_jc_j+|C_j|^2C_j\nonumber\\
    &+&|C_{j-1}|^2C_{j-1}+|C_{j+1}|^2C_{j+1}=0\, ,
    \label{supermode3Delta}
\end{eqnarray}
where $C_{j}\equiv c_{j+1}+c_{j}+c_{j-1}$.
It is now evident that the degree of nonlocality has increased (thus inducing a long-range nonlinear wave mixing process) relative to the previous cases. That is, the nonlinear coupling goes beyond nearest neighbors and includes optical fields located at sites $j\pm 2$ (on top of $j$, $j\pm1$). As a result, one might expect a fundamental difference in the statistical behavior of the projection coefficients. We have numerically solved Eq.~\eqref{supermode3Delta} under the same computational conditions as in the $N=2$ case and despite the sparsity of the nonlocal mixing terms the statistical distribution of the modal amplitudes $\langle|c_j|^2\rangle$ almost approached the theoretically predicted RJ distribution as is seen in Fig.~\ref{All_Delta} (b). To further clarify the role that nonlinear nonlocal wave mixing plays in accelerating the process of reaching the thermalization state, we have simulated Eq.~\eqref{SupermodeDeltaDynamics} for $N=4$ and $N=5$. The results were in good agreement with the theoretically predicted RJ distribution (see Fig.~\ref{All_Delta} (c) and (d)) .
\end{itemize}
\begin{figure*}[ht]
    \centering
    \includegraphics[width=\linewidth]{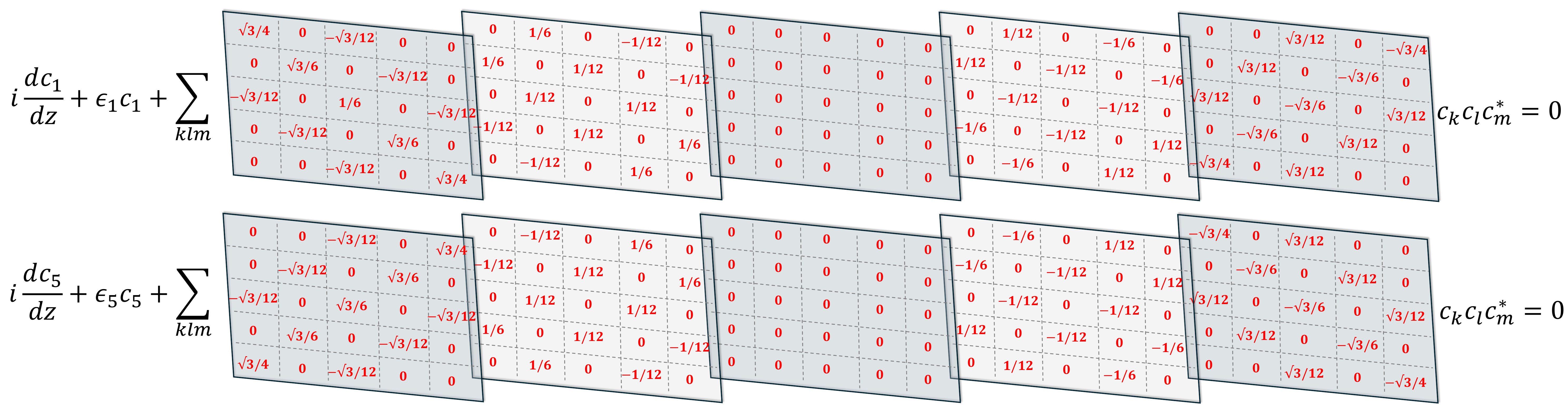}
    \caption{Evolution equations of the projection coefficients $c_j$ with $j=1,5$ corresponding to the Ablowitz-Ladik model given by Eq.~\eqref{SupermodeAL} when the number of supermodes is $M=5$. The index labeling is identical to those in Fig.~\ref{TensorKerr_M5} .}
    \label{TensorAblad}
\end{figure*}
\section{\label{sec7}Lattices with broken tensorial symmetries} 
In this section, we present examples whereby cubic nonlinear optical lattices with broken quasi-Hermiticity and permutation symmetries fail to thermalize to a Rayleigh-Jeans distribution. We first investigate an integrable system, namely the Ablowitz-Ladik (AL) model \cite{ablowitzLadik1976, ablowitz1991solitons} which falls under the category of cubic lattices that remain invariant under the gauge transformation given in Eq.~\eqref{gauge}. The main reason behind this choice is the fact that the AL lattice does not thermalize \cite{SelimPyrialakos_Ablowitz}, as such, it would be interesting to see if the associated tensorial symmetries are broken or not. Interestingly enough, as we shall see later, none of the conserved quantities of the AL model represents the conventional power given by Eq.~\eqref{Power} and the lattice cannot be derived from a Hamiltonian using the standard Poisson brackets introduced in Sec.~\ref{sec2}. Consequently, we aim to verify that the mixing tensor of the governing equation for this integrable system does not remain invariant under the previously established symmetries and compare it to our previous cases. This comparative analysis will deepen our understanding of thermalization from the first principles by identifying the discrepancies of similarly structured lattices. The other example that will be used to test our possible symmetry-thermalization connection is a lattice that conserves only power.

\begin{figure}[!hbt]
    \centering
    \includegraphics[width=\linewidth]{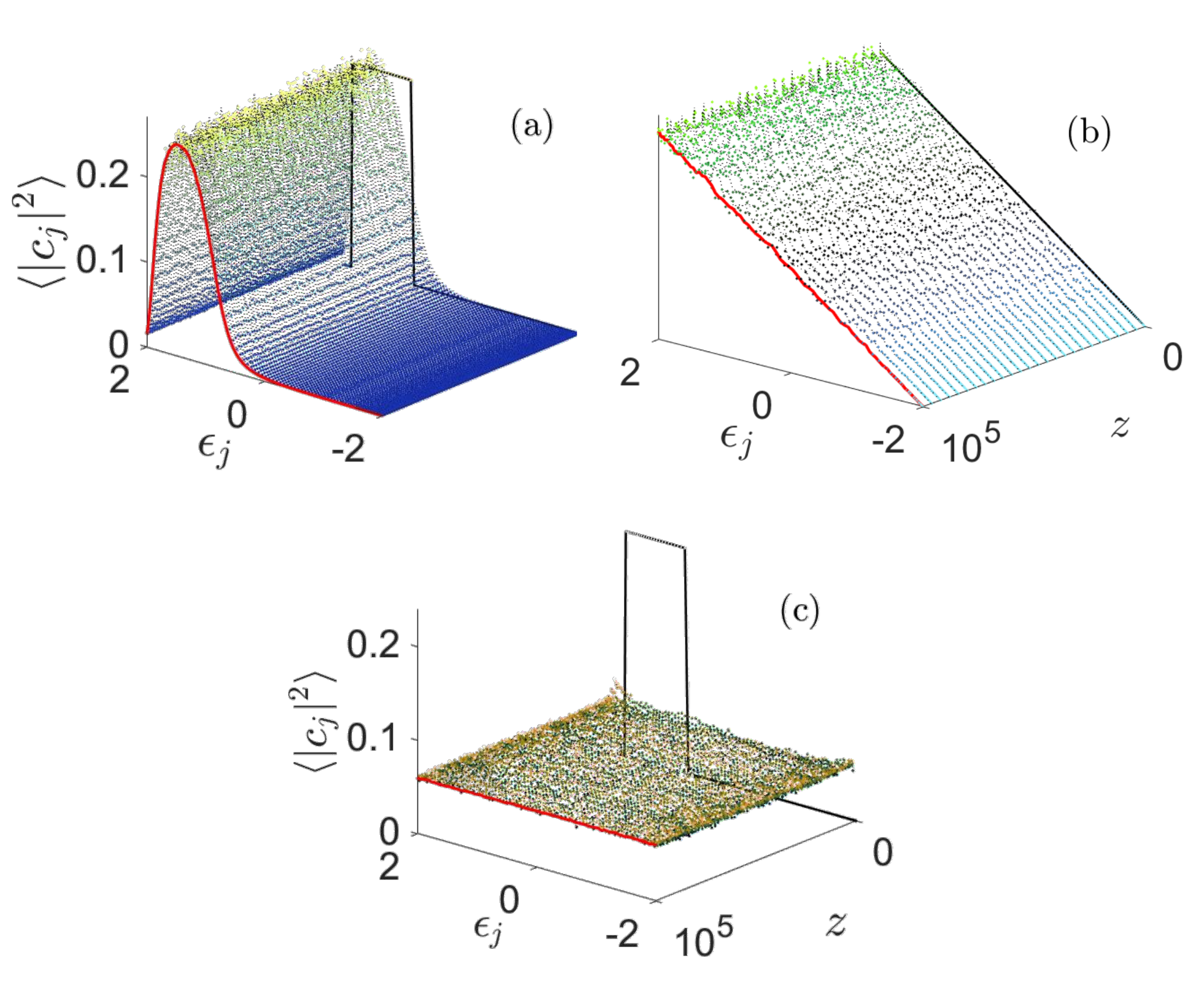}
    \caption{(a),(b) Dynamic evolution of the Ablowitz-Ladik lattice and (c) the nonlocal lattice that conserves only power, given by Eq.~\eqref{OnlyPowerLattice2}. The total power of each system is fixed at $P=6$. At $z=0$ (solid black line), power is equally distributed among 25 modes in the range $0<\epsilon_j<2$ in (a) and (c) and linearly distributed among all modes in (b). Averaging over 800 realizations with randomly perturbed phase initial conditions, and after a propagation distance of $z=100000$, none of the systems converge to a RJ distribution. The AL lattice equilibrates to different distributions (red solid lines) depending on the initial power arrangement in (a) and (b), whereas the lattice that conserves only power attains equipartition (c).}
    \label{All_NonThermalization}
\end{figure}

We begin with the AL model defined on the finite set of integers $n=1,2,...,M$ given by
\begin{equation}
    i\frac{dA_n}{dz}+A_{n+1}+A_{n-1}+|A_n|^2(A_{n-1}+A_{n+1})=0\,,
    \label{AblowitzLadik}
\end{equation}
which is known to be an integrable model possessing $M$ number of conservation laws. The first two are
\begin{equation}
    P_{\text{AL}} =\sum_{n=1}^M\ln(1+|A_n|^2)\,,
    \label{AL_Power}
\end{equation}
and
\begin{equation}
H_{\text{AL}}=\sum_{n=1}^M A_n^*A_{n+1} + A_{n}A_{n+1}^*  \, .
\label{AL_Hamiltonian}
\end{equation}
It is important to note that the AL model can be derived from the above Hamiltonian using 
\begin{equation}
    \frac{dA_n}{dz}=i\{A_n,H_{\text{AL}}\}\, ,
    \label{HamiltonEq_AL}
\end{equation}
where now $\{,\}$ denotes the non-standard Poisson bracket defined by
\begin{equation}
    \{D,\tilde D\}=\sum_{j=1}^M\left(\frac{\partial D}{\partial A_j}\frac{\partial \tilde D}{\partial A_j^*}-\frac{\partial D}{\partial A_j^*}\frac{\partial \tilde D}{\partial A_j}\right)\left(1+|A_j|^2\right)\, ,
\end{equation}
where $D$ and $\tilde D$ are arbitrary functionals of the canonical variables $A_n$, $A^*_n$. 
The AL equation in modal base assumes the form
\begin{equation}
    i\frac{dc_j}{dz}+c_j\epsilon_j+\sum\limits_{klm} T^{AL}_{j,k,l,m}c_k c_l c^*_m=0\, ,
    \label{SupermodeAL}
\end{equation}
where the Ablowitz-Ladik tensor is given as
\begin{equation}
    T^{AL}_{j,k,l,m}\! = \!B
    \sum\limits_{n=1}^{M}\!\big(\!\sin(l x_{n+1})+\sin(l x_{n-1})\big)\!\prod_{i=1}^3 \sin(p_i x_{n}) ,
    \label{TensorAL}
\end{equation} 
here, $\{p_i\}_{i=1}^3 \equiv \{j,m,k \}$. A simple example that illustrates the algebraic structure of Eq.~\eqref{SupermodeAL} is demonstrated in Fig.~\ref{TensorAblad}. An important and immediate result is that the tensor $T^{AL}_{j,k,l,m}$ breaks the quasi-Hermiticity (invariance under $k\leftrightarrow l$ and $j\leftrightarrow m$) and permutation symmetries (invariance under $k\leftrightarrow m$ and $l\leftrightarrow j$). As a result, the equilibrium state of such a lattice system, if it exists, should not conform to a RJ distribution according to our tensorial symmetries-thermalization proposition. This assertion is substantiated by direct numerical simulations as shown in Fig.~\ref{All_NonThermalization} (a) and (b) (see also \cite{SelimPyrialakos_Ablowitz}).
It is important at this point to emphasize that since the symmetries (Eq.~\eqref{QuasiHermiticity} and Eq.~\eqref{Permutation}) are not preserved, one cannot derive Eq.~\eqref{SupermodeAL} from the Hamilton's equation of motion using the Hamiltonian given Eq.~\eqref{NonLinearHamSupermode} with $T_{j,k,l,m}=T^{AL}_{j,k,l,m}$ as presented in Sec.~\ref{sec2_3}.  
It can be shown that the mixing tensor of the Ablowitz-Ladik lattice is intrinsically related to that of the Kerr via
\begin{equation}
    T^{\text{AL}}_{j,k,l,m} =2T^{\text{Kerr}}_{j,k,l,m}\cos(kx_n)\, .
\end{equation}
Notice that if $M$ is even $\cos(kx_n)$ is never zero. Consequently, the locations of the non-zero entries of the Kerr and Ablowitz-Ladik tensors are identical. This observation highlights the significance of preserving or breaking the tensorial quasi-Hermiticity and permutation symmetries due to the totally different thermalization behavior of the two lattices. This in itself indicates that knowledge of the locations of the non-zero elements of the tensor is not sufficient to guarantee thermalization. It should be emphasized that thermalization (or lack thereof) is independent of whether the number of modes is odd or even, making these extra zeros that will emerge when $M$ is odd inconsequential to the equilibrium distribution, compared to the effect that symmetry breaking has.

We next provide a case that supports our symmetry preserving-thermalization link. In previous sections (\ref{sec5}, \ref{sec6}) we discussed several circumstances whereby lattices with two conservation laws (that respect the tensorial symmetries) lead to a RJ distribution. Here, we show how breaking the permutation symmetries does not give rise to an equilibrium state that follows the Rayleigh-Jeans law. For that purpose, we will consider a lattice that solely conserves the power but cannot necessarily be derived from a real Hamiltonian function via Hamilton's equations of motion. As a result, the only symmetry that is preserved is the quasi-Hermiticity ($k\leftrightarrow m$ and $l\leftrightarrow j$). If a thermal equilibrium state exists, then its distribution will follow an equipartition of power. More precisely, we consider the following cubic lattice:
\begin{equation}
    i\frac{dA_n}{dz}+A_{n+1}+A_{n-1}+|A_{n+1}|^2A_{n}=0\,.
    \label{OnlyPowerLattice2}
\end{equation}
In supermode base, the above equation takes the form
\begin{equation}
    i\frac{dc_j}{dz}+c_j\epsilon_j+\sum\limits_{k,l,m,=1}^M T_{j,k,l,m}c_k c_l c^*_m=0\,,
    \label{SupermodeOnlyPowerLat1}
\end{equation}
where the corresponding mixing tensor is given by
\begin{eqnarray}
    T_{j,k,l,m} &=& B
    \sum\limits_{n=1}^{M} \prod\limits_{i=1}^2  \sin(q_i x_{n})\sin(q_{i+2} x_{n+1})\, .
    \label{TensorOnlyPower2}
\end{eqnarray} 
It can be shown that the above tensor remains invariant under the index change $k\leftrightarrow m$ and $l\leftrightarrow j$. However, the permutation symmetries $k\leftrightarrow l$ and $j\leftrightarrow m$ are broken. This comes as no surprise since, by construction, the given system is not even Hamiltonian. Consequently, the system will always thermalize to equipartition as the numerical simulations indicate (see Fig.~\ref{All_NonThermalization} (c)).


\section{\label{sec9}Conclusion}
In this paper, we have attempted to establish a connection between thermalization properties of cubic nonlinear optical lattices and their tensorial symmetries that arise from conservation of power and Hamiltonian. We have provided several examples of cubic nonlinear lattices for which we unlocked the internal structure of their nonlinear mixing tensors which also preserved the above-mentioned symmetries and have been shown to equilibrate to a RJ distribution. In addition, an inverse approach is developed whereby one departs from an evolution equation in supermode base with short range nonlinear mixing terms and arrives at its counterpart in local base. Numerical simulation of such novel systems reveals the connection between the degree of nonlinear wave mixing and thermalization. Lastly, we presented examples of cubic nonlinear lattices with broken quasi-Hermiticity and permutation symmetries whose equilibrium state does not conform with a RJ distribution. In general, these findings underscore the importance of the internal structure of the mixing tensor in governing thermalization and open new avenues for exploring the interaction between nonlinearity, nonlinear symmetries, and thermalization in complex lattice systems.

\section{Acknowledgments}
K.G.M. work was funded by the European Research Council (ERC-Consolidator) under grant agreement No. 101045135 (Beyond Anderson).
D.N.C. work was partially supported by the Army Research Office (W911NF-23-1-0312), the MPS Simons collaboration (Simons grant no. 733682), by the Air Force Office of Scientific Research (AFOSR) Multidisciplinary University Research Initiative (MURI) award on Novel light-matter interactions in topologically non-trivial Weyl semimetal structures and systems (award no. FA9550-20-1-0322), AFOSR MURI award on Programmable systems with non-Hermitian quantum dynamics (award no. FA9550-21-1-0202), the Department of Energy (DE-SC0022282), W.M. Keck Foundation, the Department of Energy (DE-SC0025224), and the Government of Israel, Ministry of Defense (4441279927).

\section*{\label{AppendA}Appendix A}
In Sec.~\ref{sec2} we introduced the main equation for cubic nonlinear lattices in the modal basis governing the nonlinear evolution of the projection coefficients $c_j(z)$ (see Eq.~\eqref{SuperModeDNLSE}). From this, one can deduce the dynamics of the modal occupancies $|c_j(z)|^2$:
\begin{equation}
    \frac{d|c_j|^2}{dz} = 2\,\text{Im}\left(\sum_{klm} T_{j,k,l,m}^*\, c_j\, c_k^*\, c_l^*\, c_m\right)\,.
    \label{ModalOccEq}
\end{equation}
This equation provides insight into the conditions required for the conservation of power $P$ as defined in Eq.~\eqref{PowerSupermode}. This in tern, imposes certain constraint on the tensor $T_{j,k,l,m}$ given by
\begin{equation}
    \text{Im}\left(F\right) = 0\,,
\end{equation}
where,
$\displaystyle F = \sum_{jklm} T_{j,k,l,m}^*\, c_j\, c_k^*\, c_l^*\, c_m$ .
Then, power conservation requires \( F \) to be real. One can show that this is the case if and only if the symmetry constraint postulated in Eq.~\eqref{QuasiHermiticity} is valid.
To see that, we consider the quantity 
\begin{align}
    F - F^* &= \sum_{jklm} T_{j,k,l,m}^*\, c_j\, c_k^*\, c_l^*\, c_m \nonumber\\
    &- \sum_{j'k'l'm'} T_{j',k',l',m'}\, c_{j'}^*\, c_{k'}\, c_{l'}\, c_{m'}^*\,.
    \label{F-F*}
\end{align}
Relabeling the indices in the second sum of Eq.~\eqref{F-F*} 
i.e., $j' = l,\,\, k' = m,\,\, l' = j,\,\, m' = k\,,$
we get
\begin{equation}
    F - F^* = \sum_{jklm} \Bigl( T_{j,k,l,m}^* - T_{l,m,j,k} \Bigr) c_j\, c_k^*\, c_l^*\, c_m\,.
\end{equation}
Therefore, if the symmetry condition given in Eq.~\eqref{QuasiHermiticity} holds, then $F$ is real. In all the cases considered in this work, the mixing tensor $T_{j,k,l,m}$ is real. Hence, the transformations $ k \longleftrightarrow m $ and $l \longleftrightarrow j $ (referred to as quasi-Hermiticity) are sufficient to ensure conservation of power.

We now proceed to examine the Hamiltonian structure introduced in Eqs.~\eqref{HamiltonEq}, \eqref{LinearHamSupermode}, and \eqref{NonLinearHamSupermode} and identify the additional restrictions required for its conservation. We begin with the Hamiltonian
\begin{align}
    \mathcal{H} =\sum\limits_{j=1}^{M}\epsilon_j|c_j|^2 &+\frac{1}{4}\sum\limits_{jklm}T_{j,k,l,m}c_jc^*_k c^*_l c_m\nonumber\\
    &+\frac{1}{4}\sum\limits_{jklm}T^*_{j,k,l,m}c^*_jc_k c_l c^*_m \,.
\end{align}
The equation of motion for the modal amplitude as given in Eq.~\eqref{HamiltonEq} becomes
\begin{equation}
    \frac{dc_j}{dz} = i \frac{\partial \mathcal{H}}{\partial c_j^*}\, ,
\end{equation}
from which we obtain
\begin{align}
    &i\frac{dc_j}{dz}=-\epsilon_jc_j -\frac{1}{4}\Big(\sum_{j'lm}T_{j'jlm} c_{j'}c^*_l c_m \nonumber\\
    &+ \sum_{j'km}T_{j'kjm} c_{j'} c^*_k c_m  + \sum_{j'kl}T^*_{j'klj} c^*_{j'} c_k c_l  \nonumber\\
    &+ \sum_{klm}T^*_{jklm} c_k c_l c^*_m \Big)\, .
\end{align}
Keep in mind that $j'$, $k$, $l$, and $m$ are indices intrinsic to the Hamiltonian, so it's impossible to group all terms under one sum. However, if we require that the indices $k$ and $l$ are interchangeable and the same for $j'$ and $m$ (i.e., $k \leftrightarrow l$ and $j' \leftrightarrow m$), then the first two sums can be merged, as can the last two. In that case, we have
\begin{align}
    i\frac{dc_j}{dz}=-\epsilon_jc_j &-\frac{1}{2}\sum_{j'km}T_{j'jkm} c_{j'} c^*_k c_m  \nonumber\\
    &-\frac{1}{2}\sum_{klm}T^*_{jklm} c_k c_l c^*_m \, .
\end{align}
Furthermore, by invoking the conservation of power (which imposes invariance under the permutation and complex conjugation of the indices $k\leftrightarrow m$ and $l \leftrightarrow j'$, we deduce that
\begin{align}
    i\frac{dc_j}{dz}&+\epsilon_jc_j +\sum_{klm}T_{jklm} c_k c_l c^*_m = 0\, .
\end{align}
This is precisely the evolution equation in the supermode basis (see Eq.~\eqref{SuperModeDNLSE}). Thus, the above analysis shows that, in order for the Hamiltonian to be conserved, the mixing tensor $T_{j,k,l,m}$ must satisfy the additional symmetry condition, namely the permutation symmetries $k\longleftrightarrow l$ and $j \longleftrightarrow m$.
\section*{\label{AppendB}Appendix B}
Starting from the tensorial definition for the nonlinear free lattice with Kerr nonlinearity as given in Eq. \eqref{TensorKerrSine}, we get, by writing the sine in its exponential form
\begin{equation}
    T_{j,k,l,m}^{\text{Kerr}}=\frac{B}{16}\sum_{\iota=1}^8(-1)^{\iota+1}\gamma^{(\iota)}_{j,k,l,m}\, ,
    \label{BIGSUM}
\end{equation}
where as a reminder, $B=4/(M+1)^2$, $x_n=n\pi/(M+1)$, $w_{\iota}$ are given by Table~\ref{Table_w} and
\begin{equation}
    \gamma^{(\iota)}_{j,k,l,m}\equiv \sum_{n=1}^M \left(e^{i w_{\iota}x_n}+e^{-iw_{\iota}x_n}\right)\, .
    \label{gamma_definition}
\end{equation}
By performing the summation over $n$ we get
\begin{align}
    \gamma_{j,k,l,m}^{(\iota)}&=2 \text{Re}\left(\frac{(-1)^{w_{\iota}}-e^{ix_{w_{\iota}}}}{e^{ix_{w_{\iota}}}-1}\right)\label{gamma_step1}\\
    \nonumber\\
    &= \frac{(-1)^{w_{\iota}+1}-1+\big((-1)^{w_{\iota}}+1\big)\cos\left(x_{w_{\iota}}\right)}{1-1\cos\left(x_{w_{\iota}}\right)}\\
    \nonumber\\
    &=(-1)^{w_{\iota}+1}-1\, .
    \label{gamma_firstvalue}
\end{align}
The above result is valid whenever $x_{w_{\iota}}\neq 2\pi \kappa_{\iota}$, with $\kappa_{\iota}$ an arbitrary integer. In the case where $x_{w_{\iota}} = 2\pi \kappa_{\iota}$ or equivalently $w_{\iota}=2(M+1)\kappa_{\iota}$, Eq.~\eqref{gamma_definition} gives
\begin{equation}
     \gamma_{j,k,l,m}^{(\iota)} =2M\, .
     \label{gamma_2M}
\end{equation}
Thus, the value of the tensor $\gamma_{j,k,l,m}^{(\iota)}$ is given by expression \eqref{gamma_2M} if $w_{\iota}=2(M+1)\kappa_{\iota}$ or by Eq.~\eqref{gamma_firstvalue} if $w_{\iota}\neq 2(M+1)\kappa_{\iota}$. To this end Eq.~\eqref{M3_j} now reads
\begin{eqnarray}
    i\frac{dc_1}{dz}&+&\left(\epsilon_1+P/2+(2|c_3|^2-|c_1|^2)/8\right)c_1\nonumber\\
    &+& \frac{1}{8}\bigg(2c_2^2c_1^*-2|c_1|^2c_3+3c_3^2c_1^*
    +4|c_2|^2c_3\nonumber\\
    &-&c_1^2c_3^*+2c_2^2c_3^*-|c_3|^2c_3\bigg)=0\, ,
    \label{M3_1_full}
\end{eqnarray}
\begin{eqnarray}
    i\frac{dc_2}{dz}&+&\left(\epsilon_2+P/2\right)c_2+\frac{1}{2}\bigg(c_2c_3c_1^*
    +c_1c_3c_2^*\nonumber\\
    &+&\frac{1}{2}c_1^2c_2^*+\frac{1}{2}c_3^2c_2^*+c_2c_1c_3^*\bigg)=0\, ,
    \label{M3_2_full}
\end{eqnarray}
\begin{eqnarray}
    i\frac{dc_3}{dz}&+&\left(\epsilon_3+P/2+(2|c_1|^2-|c_3|^2)/8\right)c_3\nonumber\\
    &+&\frac{1}{8}\bigg(2c_2^2c_1^*-|c_1|^2c_1-c_3^2c_1^*+4|c_2|^2c_1\nonumber\\
    &+&3 c_1^2c_3^*- 2|c_3|^2c_1+2c_2^2c_3^*\bigg)=0\, ,
     \label{M3_3_full}
\end{eqnarray}
from which one can identify the non-mixing ${\cal N}_j$ and the mixing ${\cal M}_j$ terms that appear in Eq.~\eqref{M3_j}.
In a similar fashion, one can derive the expressions for the auxiliary tensor $\zeta^{(\iota)}_{j,k,l,m}$ that appear in Eq.~\eqref{Xi_Tensor}. Starting form the definition
\begin{equation}
    \zeta_{j,k,l,m}^{(\iota)} = \frac{1}{2i} \sum\limits_{n=1}^M\Big(e^{i w_{\iota}x_n}-e^{-iw_{\iota}x_n}\Big)\, ,
\end{equation}
we have 
\begin{align}
    \zeta_{j,k,l,m}^{(\iota)}&= \text{Im}\left(\frac{(-1)^{w_{\iota}}-e^{ix_{w_{\iota}}}}{e^{ix_{w_{\iota}}}-1}\right)\label{zeta_step1}\\
    \nonumber\\
    &= \frac{\left((-1)^{w_{\iota}+1}+1\right)\sin\left(x_{w_{\iota}}\right)}{1-\cos\left(x_{w_{\iota}}\right)}\\
    \nonumber\\
    &=\frac{1-(-1)^{w_{\iota}}}{2}\cot\left(x_{w_{\iota}}\right)\, .
    \label{zeta_laststep}
\end{align}
As before, the result in Eq.~\eqref{zeta_laststep} is valid when $w_{\iota}\neq2\pi\kappa_{\iota}$. In the case where $\displaystyle w_{\iota}=2\pi\kappa_{\iota}$, we get
\begin{equation}
    \zeta_{j,k,l,m}^{(\iota)}=0\, .
\end{equation} 

\nocite{*}

\bibliography{bibliography}

\end{document}